\documentclass[aps,prc,twocolumn,twoside,showkeys,showpacs,floatfix]{revtex4}
\usepackage{graphics,epsfig}

\newcommand{\figart}[5]{\begin{figure}[#1]
			\begin{center}
                        \includegraphics*[scale=#5,draft=false]{./#2.eps}
                        \caption{#3}
                        \label{#4}
                        \end{center}
                        \end{figure}}

\newcommand{\ctp}[0]{$\cos(\theta_{prox})$}
\newcommand{\ctfqp}[0]{$\cos(\theta_{F/QP})$}
\newcommand{\Vz}[0]{$V_{z}$}
\newcommand{\Vr}[0]{$V_{rel}$}
\newcommand{\Vcm}[0]{$V_{c.m.}$}
\newcommand{\Mfra}[0]{$M_{imf}$}
\newcommand{\Etl}[0]{$E_{t12}$}
\newcommand{\AMeV}[0]{$\rm{A}\,\rm{MeV}$}

\newcommand{\mod}[1]{{#1}}
\newcommand{\modi}[1]{{#1}}
\newcommand{\modif}[1]{{#1}}

\newcommand{\LPC}[0]{\it LPC Caen (IN2P3-CNRS/ENSICAEN et Universit\'e),
F-14050 Caen Cedex , France}

\begin{document}

\title{\large \bf Dynamical effects in multifragmentation
at intermediate energies.\thanks{Experiment performed at GANIL}}

\author{J.~Colin}
\affiliation{\LPC}
\author{D.~Cussol}
\affiliation{\LPC}
\author{J.~Normand}
\affiliation{\LPC}
\author{N.~Bellaize}
\affiliation{\LPC}
\author{B.~Borderie}
\affiliation{ Institut de Physique Nucl\'eaire, IN2P3-CNRS, F-91406 
Orsay Cedex,  France.}
\author{R.~Bougault}
\affiliation{\LPC}
\author{B.~Bouriquet}
\affiliation{ GANIL, CEA et IN2P3-CNRS, B.P.~5027, F-14076 Caen Cedex, France.}
\author{A.M.~Buta}
\affiliation{\LPC}
\author{J.L.~Charvet}
\affiliation{ DAPNIA/SPhN, CEA/Saclay, F-91191 Gif sur Yvette Cedex, France.}
\author{A.~Chbihi}
\affiliation{ GANIL, CEA et IN2P3-CNRS, B.P.~5027, F-14076 Caen Cedex, France.}
\author{R.~Dayras}
\affiliation{ DAPNIA/SPhN, CEA/Saclay, F-91191 Gif sur Yvette Cedex, France.}
\author{D.~Durand}
\affiliation{\LPC}
\author{J.D.~Frankland}
\affiliation{ GANIL, CEA et IN2P3-CNRS, B.P.~5027, F-14076 Caen Cedex, France.}
\author{E.~Galichet}
\affiliation{ Institut de Physique Nucl\'eaire, IN2P3-CNRS, F-91406 Orsay Cedex,
 France.}
\affiliation{ Conservatoire National des Arts et M\'etiers, F-75141 Paris
Cedex 03.}
\author{D.~Guinet}
\affiliation{ Institut de Physique Nucl\'eaire, IN2P3-CNRS et Universit\'e
F-69622 Villeurbanne, France.}
\author{B.~Guiot}
\affiliation{ GANIL, CEA et IN2P3-CNRS, B.P.~5027, F-14076 Caen Cedex, France.}
\author{S.~Hudan}
\affiliation{ GANIL, CEA et IN2P3-CNRS, B.P.~5027, F-14076 Caen Cedex, France.}
\author{P.~Lautesse}
\affiliation{ Institut de Physique Nucl\'eaire, IN2P3-CNRS et Universit\'e
F-69622 Villeurbanne, France.}
\author{F.~Lavaud}
\affiliation{ Institut de Physique Nucl\'eaire, IN2P3-CNRS, F-91406 Orsay Cedex,
 France.}
\author{N.~Le~Neindre}
\affiliation{ Institut de Physique Nucl\'eaire, IN2P3-CNRS, F-91406 Orsay Cedex,
 France.}
\author{O.~Lopez}
\affiliation{\LPC}
\author{L.~Manduci}
\affiliation{\LPC}
\author{J.~Marie}
\affiliation{\LPC}
\author{L.~Nalpas}
\affiliation{ DAPNIA/SPhN, CEA/Saclay, F-91191 Gif sur Yvette Cedex, France.}
\author{M.~P\^arlog}
\affiliation{ National Institute for Physics and Nuclear Engineering, RO-76900
Bucharest-M\u{a}gurele, Romania.}
\author{P.~Paw{\l}owski}
\affiliation{ Institut de Physique Nucl\'eaire, IN2P3-CNRS, F-91406 Orsay Cedex,
 France.}
\author{E.~Plagnol}
\affiliation{ Institut de Physique Nucl\'eaire, IN2P3-CNRS, F-91406 Orsay Cedex,
 France.}
\author{M.~F.~Rivet} 
\affiliation{ Institut de Physique Nucl\'eaire, IN2P3-CNRS, F-91406 
Orsay Cedex,  France.} 
\author{E.~Rosato}
\affiliation{ Dipartimento di Scienze Fisiche e Sezione INFN, Universit\`a di 
Napoli ``Federico II'', I-80126 Napoli, Italy.}
\author{R.~Roy}
\affiliation{ Laboratoire de Physique Nucl\'eaire, Universit\'e Laval,
Qu\'ebec, Canada.}
\author{J.C.~Steckmeyer}
\affiliation{\LPC}
\author{B.~Tamain}
\affiliation{\LPC}
\author{A.~Van Lauwe}
\affiliation{\LPC}
\author{E.~Vient}
\affiliation{\LPC}
\author{M.~Vigilante}
\affiliation{ Dipartimento di Scienze Fisiche e Sezione INFN, Universit\`a di 
Napoli ``Federico II'', I-80126 Napoli, Italy.}
\author{C.~Volant}
\affiliation{ DAPNIA/SPhN, CEA/Saclay, F-91191 Gif sur Yvette Cedex, France.}
\author{J.P.~Wieleczko}
\affiliation{ GANIL, CEA et IN2P3-CNRS, B.P.~5027, F-14076 Caen Cedex, France.}
\author{(INDRA Collaboration)}
\noaffiliation

\begin{abstract} 

The fragmentation of the quasi-projectile is \mod{studied} with the INDRA
multidetector for different colliding systems \mod{and} incident energies
in the Fermi energy range. Different experimental observations show that a
large part of the fragmentation is not compatible with the statistical
fragmentation of a fully \modi{equilibrated} nucleus. The study of internal
correlations is a powerful tool, especially to evidence entrance channel 
effects. These effects have to be included in the theoretical descriptions of
nuclear multifragmentation. 

\end{abstract}

\pacs{25.70.-z,25.70.Mn,25.70.Pq}
\keywords{nucleus-nucleus collisions, multifragmentation, dynamical effects}

\maketitle

\section*{Introduction.}

 The physical
characteristics (shapes, excitation energies, angular momentum, radial flow,
...) of the \modi{hot} nuclei \modi{formed in nucleus-nucleus collisions} 
are obtained by comparisons between models and data.
Statistical decay models are widely used and dedicated to describe the decay of
hot \mod{fully} equilibrated systems, \mod{defined as systems having reached 
energy, shape and isospin equilibrium }
\cite{Weis37,Bohr39,SIMON,SMM,MMMC,Moretto95,Moretto97,Tso95,Charity88}. 
However other experimental works show strong
effects of the entrance channel on the decay modes of the 
\modi{formed hot} systems
\cite{Stu92,Lec95,Plagnol2000,Bocage2000,Davin2002,Gingras2002,Lefort2000,Dore2000b,Cas93,Ste95,Tok95,Mon94,Glassel83,Lar97}. 
These effects \modi{which} are not taken into account in
the statistical \modi{decay models can provide} complementary
information on the nuclei such as the characteristics of the nucleon-nucleon
interaction \cite{Cussol2002,West98,Reis98,Gut90} and/or the viscosity of the 
nuclear matter.

Up to now, these entrance channel effects have been studied by tracking
deviations from the standard fission \mod{process} 
in the quasi-projectile break-up 
\cite{Bocage2000,Davin2002,Ste95,Cas93,Glassel83}. 
Such \modi{deviations} have been seen on the
angular distributions, which show the \mod{focusing} of the break-up axis along
the quasi-projectile velocity direction. \modif{Associated with} this \mod{focusing}, 
highly asymmetrical break-ups have also been observed,  whereas symmetrical
break-ups were expected. The corresponding relative velocities were higher than
the value predicted by the Viola systematics \cite{Viola}. These aligned
break-ups represent up to 75\% of the binary break-up for the Xe+Sn system
\cite{Bocage2000}.  In this paper, we will extend these studies performed for
binary break-up  to higher fragment multiplicities to 
\modi{test} to \modif{what} extent
these effects are present for other exit channels. \modi{We will present
systematic studies on
experimental data only, over a wide range of system sizes, 
incident energies and projectile/target asymmetries. There will be no 
attempt to compare these data to theoretical calculations.}

We have studied the decay of different \modi{quasi-projectiles}
\mod{formed in several projectile-target collisions} 
from 25 to 90 \AMeV. In the first section, we
present the experimental set-up, the studied sytems and the
selections which have been applied on the data. In
the second section the charge distributions, the associated velocity and
angular distributions obtained for the different systems are \modi{shown}. The
third section is devoted to the evolutions of some experimental observables
as a function of the break-up asymmetry of the \mod{quasi-projectile}.
 In the last section our conclusions are \modi{given}.

\section{The experimental set-up}
The experiments were performed at the GANIL facility with the INDRA 
detector. 
The different systems which will be presented in this article are Ni+Ni at 52
and 90 \AMeV, Xe+Sn from 39 to 50 \AMeV, Ta+Au at 33 and 39 \AMeV,  Ta+U at 33
and 39 \AMeV~and U+U at 24 \AMeV.
Target thicknesses were respectively 193 
$\rm{{\mu g/cm^2 ~of~Au}}$ for \mod{the  
experiment with the Au target}, 179 
$\rm{{\mu g/cm^2 ~of~Ni}}$ for the Ni + Ni 
experiment and 330 
$\rm{{\mu g/cm^2 ~of~Sn}}$ for the Xe + Sn 
experiment.
The \modi{100 $\rm{{\mu g/cm^2}}$ }\mod{uranium} target was deposited \mod{between two 20 $\rm{{\mu g/cm^2}}$
carbon foils} . We will show afterwards 
the method used
to separate the events \mod{corresponding} to the uranium target from those
corresponding to the \modi{carbon backing}.
Typical beam intensities were 3-4 ${\times 10^7}$ pps. 
\modi{Events} were 
registered when at least 
\mod{four} charged particle detectors fired 
(eight for the collisions with the Ta projectile). 

The INDRA detector \cite{INDRA1,INDRA2} 
can be schematically described as a set of 17 detection 
rings centered on the beam axis. In each ring the detection of charged 
products was provided with two or three detection layers. The most forward 
ring, ${2^{\circ} \le \theta_{lab} \le 3^{\circ}}$, is made of phoswich 
detectors (plastic scintillators NE102 + NE115). Between ${3^{\circ}}$ 
and ${45^{\circ}}$ 
eight rings are constituted by three detector layers: ionization chambers, 
silicon and ICs(Tl). Beyond ${45^{\circ}}$, the eight remaining 
rings are made of double layers: ionization chambers and ICs(Tl). 
The total number of 
detection cells is 336 and the overall geometrical 
efficiency of INDRA detector corresponds to 90\% of 4${\pi}$.  
Isotopic separation \modi{is} achieved up to 
Z=3-4 in the 
last layer (ICs(Tl)) over the whole angular range (${3^{\circ} \le 
\theta_{lab} \le 176^{\circ}}$). \mod{The charge resolution 
in the forward region (${3^{\circ}} \le \theta_{lab} \le 
\rm{45^{\circ}}$) \modi{is} one unit up 
to  Z$\approx$50. In the backward region 
(${\theta_{lab} \ge 45^{\circ}}$) \modi{a charge resolution of one unit
is} obtained up to  Z$\approx$20} . The energy resolution is about
5\% for ICs(Tl) and ionization chambers and better than 2\% for
Silicon {detectors}.

\figart{htb}{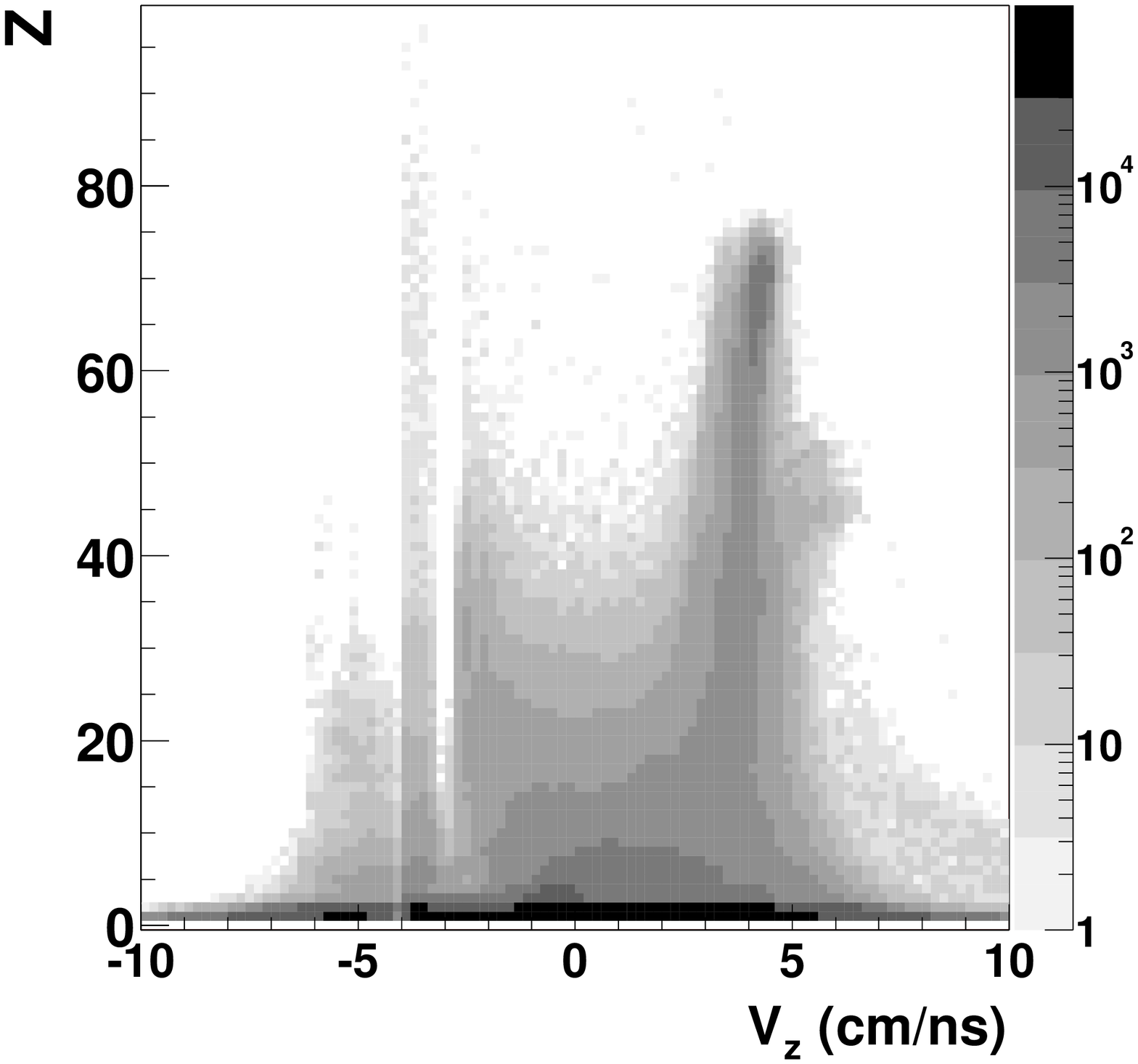}{Correlation between the charge $Z$ and the velocity
component parallel to the beam \Vz~ in the center of mass frame for the Ta+Au
collisions at 39.6 \AMeV~and for events having a total linear momentum 
greater than 80\% of the incident linear momentum. 
The shading of gray is darker for the high cross
sections.}
{f:ZV}{0.4}

\modi{The binary character of the collisions is clearly seen on figure
\ref{f:ZV} which plots the correlation between the charge $Z$ and the velocity
component parallel to the beam \Vz~of each detected fragment. The fragments are
roughly distributed around two areas: one around the quasi-projectile velocity
and charge ($V_{z}\approx 4 cm/ns$ and $Z\approx 75$) and one around the
quasi-target velocity ($V_{z}\approx -4 cm/ns$). 
Due to the experimental thresholds, the fragments emitted by the quasi-target
are not detected efficiently.}
 Then, \mod{for this analysis,} 
only the fragments produced by the
quasi-projectile are taken into account. \mod{Due to the quasi symmetry of most
of the studied systems, a} fragment \mod{(Z$\ge$3)} is assumed to be emitted by
the quasi-projectile if \modi{its velocity component parallel to the beam axis} 
is higher than the \modi{velocity of the
center of mass $V_{c.m.}$}.

In order to have enough information to perform our studies, only events for
which more than 80\% of the incident linear momentum \mod{was detected} 
are selected ($\sum Z_{i} V_{i}^{z} >
0.8 Z_{proj} V_{proj}^{z}$
where $Z_{i}$ and $Z_{proj}$ are the charge of the fragment $i$ and the charge
of the
projectile respectively and $V_{i}^{z}$ and $V_{proj}^{z}$ 
their velocity components parallel to the beam axis in the laboratory frame). 

The events \mod{are} sorted according to the multiplicity  \Mfra~
of fragments \mod{(Z$\ge$3)} emitted by the quasi-projectile. 


The use of a $4\pi$ multi-detector \modi{gives} a very good coverage of the
quasi-projectile emissions. The \modi{angular, charge and velocity or energy 
distributions are precisely obtained}. This allows
\modif{the study of} the internal correlations in one event like the correlation between
the fragments relative velocities
and the break-up direction. It will be shown later that this
is a powerful tool to distinguish different reaction mechanisms.


As said previously, the uranium was \mod{deposited between two carbon layers}.
Thus, for the systems with this target, there will be a mixing 
\mod{of reactions on}  the
\modi{backing} (C) and \mod{on} the \modi{target} (U). The separation between 
these two \mod{reactions}
is done by using the angular distribution of the quasi-projectile. For
a fixed incident energy, the grazing angle for the heavy target is 
\mod{larger} that
the grazing angle for the light one. The values of this angle are 
summarized in table \ref{t:grazing}. 

\begin{table}[htbp]
\begin{center}
\begin{tabular}{c c c}
~System~&~Incident Energy ~& ~$\theta_{grazing}~$ \\
&~(\AMeV)~& ~(degrees)~ \\
\hline
\hline
U + U & 24	& 8 \\
U + C & 24	& 0.7 \\
\hline
Ta + U & 33	& 6 \\
Ta + Au & 33	& 5 \\
Ta + C & 33	& 0.5 \\
\hline
Ta + U & 39.6	& 5 \\
Ta + Au &39.6	& 4 \\
Ta + C & 39.6	& 0.4 \\
\end{tabular}
\end{center}
\caption{Grazing angles in the laboratory frame for different systems.}
\label{t:grazing}
\end{table}

Two contributions are then expected for the \modif{quasi-projectile  
diffusion angle
$\theta_{QP}$}. The quasi-projectile \modif{(QP)} is reconstructed with the fragments whose
velocities are greater than $V_{c.m.}$. Such distributions are shown in 
\mod{figures \ref{f:tqp_tau33_all} to
\ref{f:tqp_uu24_all}}. \modi{In the first two} figures, the dashed line 
corresponds to
the system with the Ta projectile and the Au target and the full line
corresponds \modi{to} the system with the Ta projectile and the \modi{U target 
and C backing}. 
\modi{Reactions on the C backing clearly show up at small angles, while the
distributions at large angles, from their similarities with those arising from
Ta+Au, can be attributed to reaction on U only. A clear selection of Ta+U
reactions is ensured by selecting events with $\theta_{QP} > 10^\circ$.}
 For the \modi{U+C,U} system (figure
\ref{f:tqp_uu24_all}), the two contributions \modi{(U target and C backing)} 
are better separated due to the
larger difference between the values of the grazing angles. The same cut
($\theta_{QP} =10^\circ$) was applied to separate the contributions from the
uranium target and the carbon \modi{backing}. For this system, the
\modi{contribution} of the \modi{uranium} target at small angles is 
weaker compared to the Ta+C,U
systems.

\figart{htb}{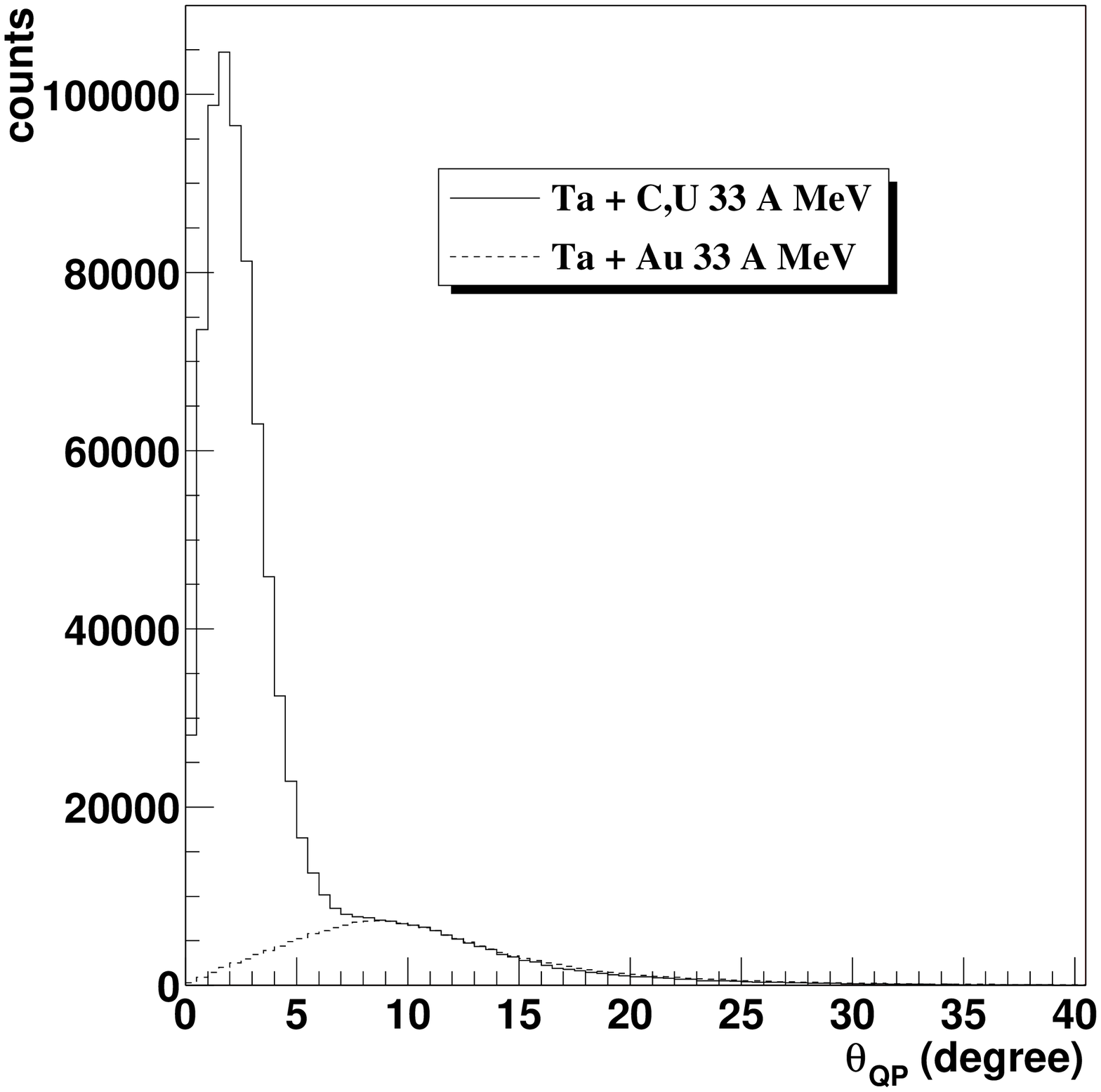}{Distribution of $\theta_{QP}$ for the Ta+C,U
system at 33 \AMeV~(full line) and the Ta+Au system at 33 \AMeV~(dashed line).}
{f:tqp_tau33_all}{0.4}
\figart{htb}{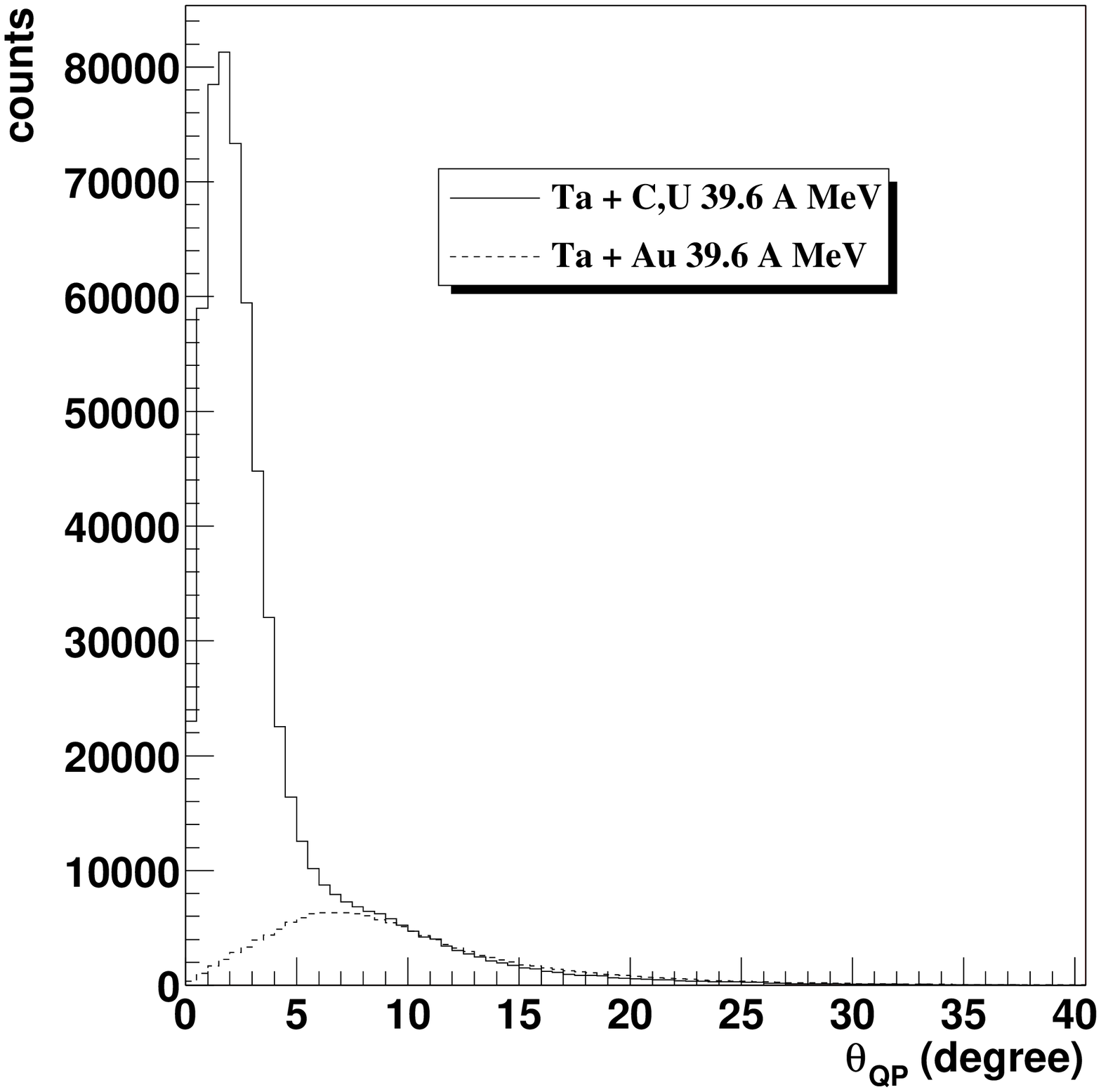}{Distribution of $\theta_{QP}$ for the Ta+C,U
system at 39.6 \AMeV~(full line) and the Ta+Au system at 39.6 \AMeV~
(dashed line).}{f:tqp_tau39_all}{0.4}
\figart{htb}{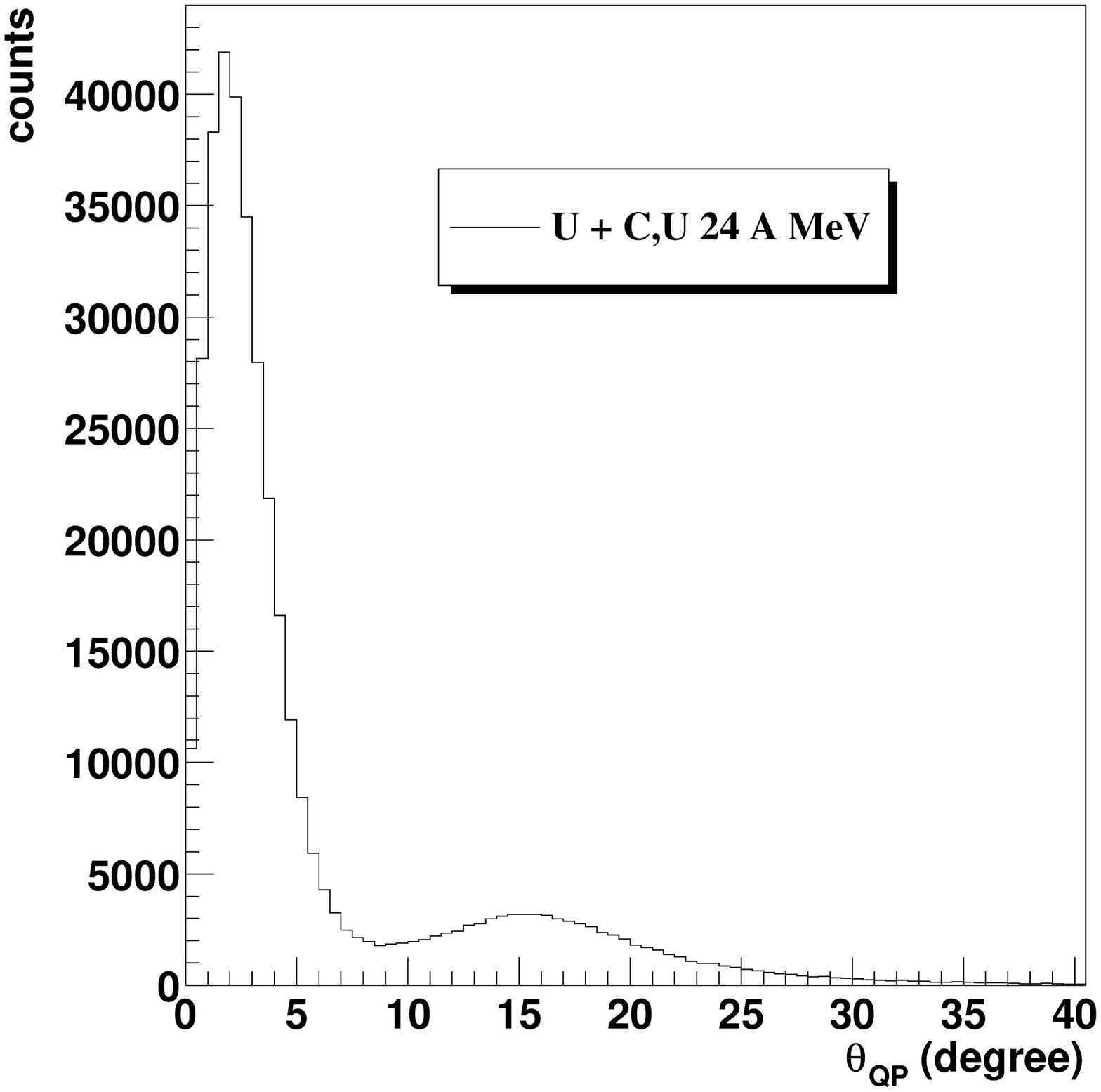}{Distribution of $\theta_{QP}$ for the U+C,U
system at 24 \AMeV.}{f:tqp_uu24_all}{0.4}
\figart{htb}{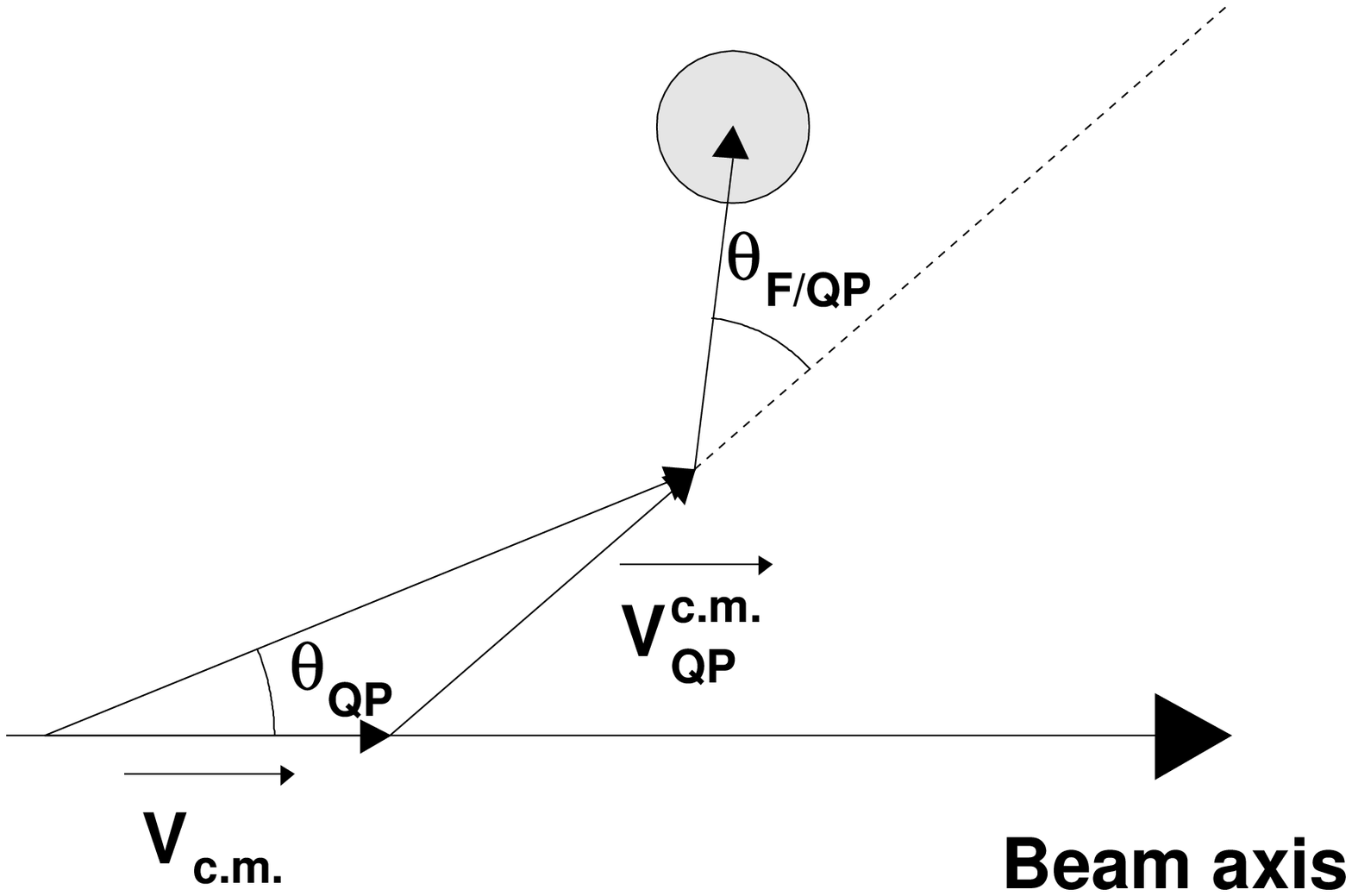}{Definition of the angles $\theta_{F/QP}$ and $\theta_{QP}$.}
{f:angles}{0.4}
\figart{htb}{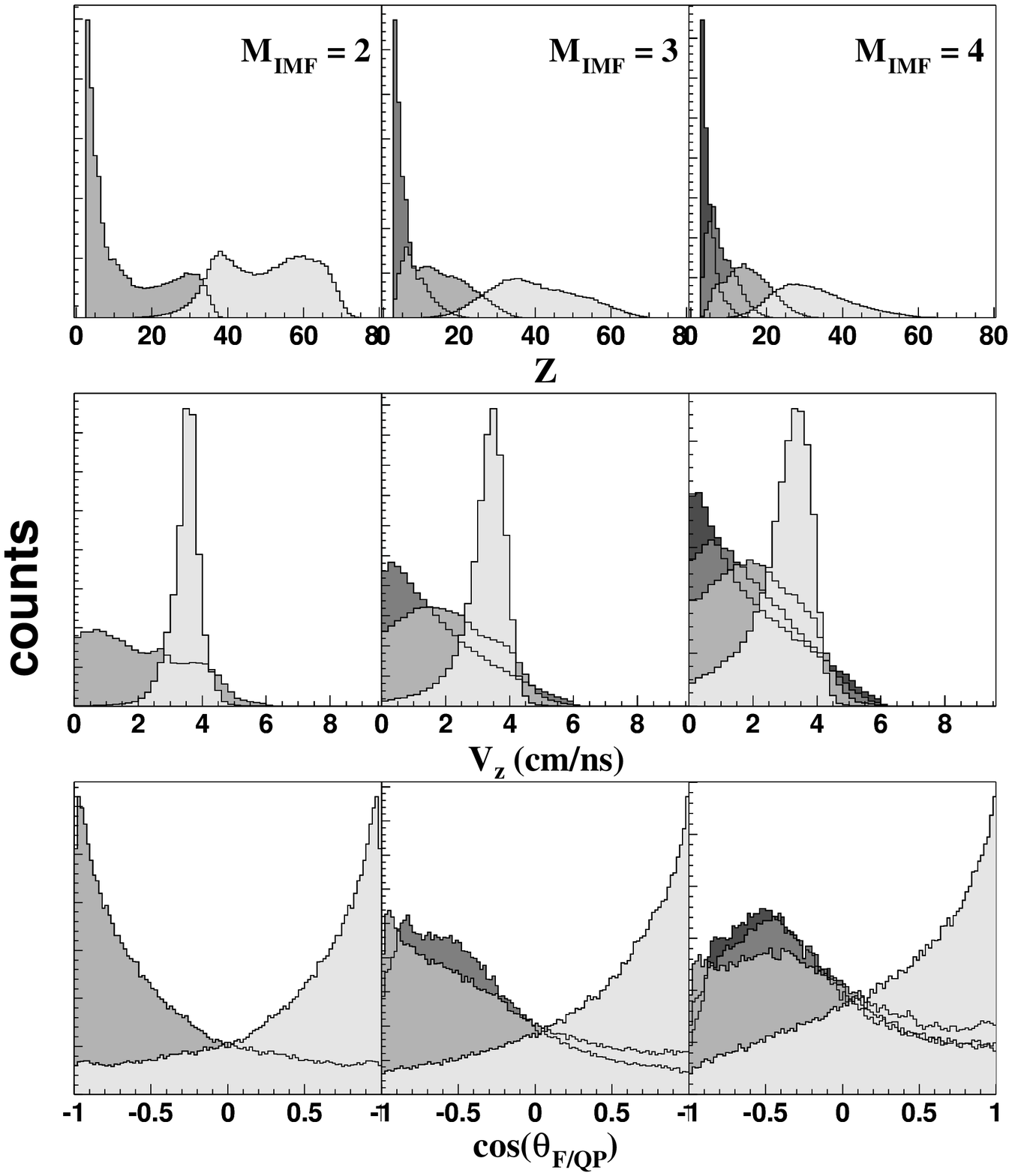}{Charge (uppermost row), parallel velocity (middle
row) and \ctfqp (lowermost row)
distributions for the Ta+Au system at 33 \AMeV. The columns correspond to
different fragment mutiplicities (from
\Mfra=2 to \Mfra=4 from left to right). The shading of the
distribution is darker and darker according to the rank of the fragment in
the event (the lightest shading correspond to the heaviest fragment in the
event).}{f:hier_taau33}{0.37}
\figart{htb}{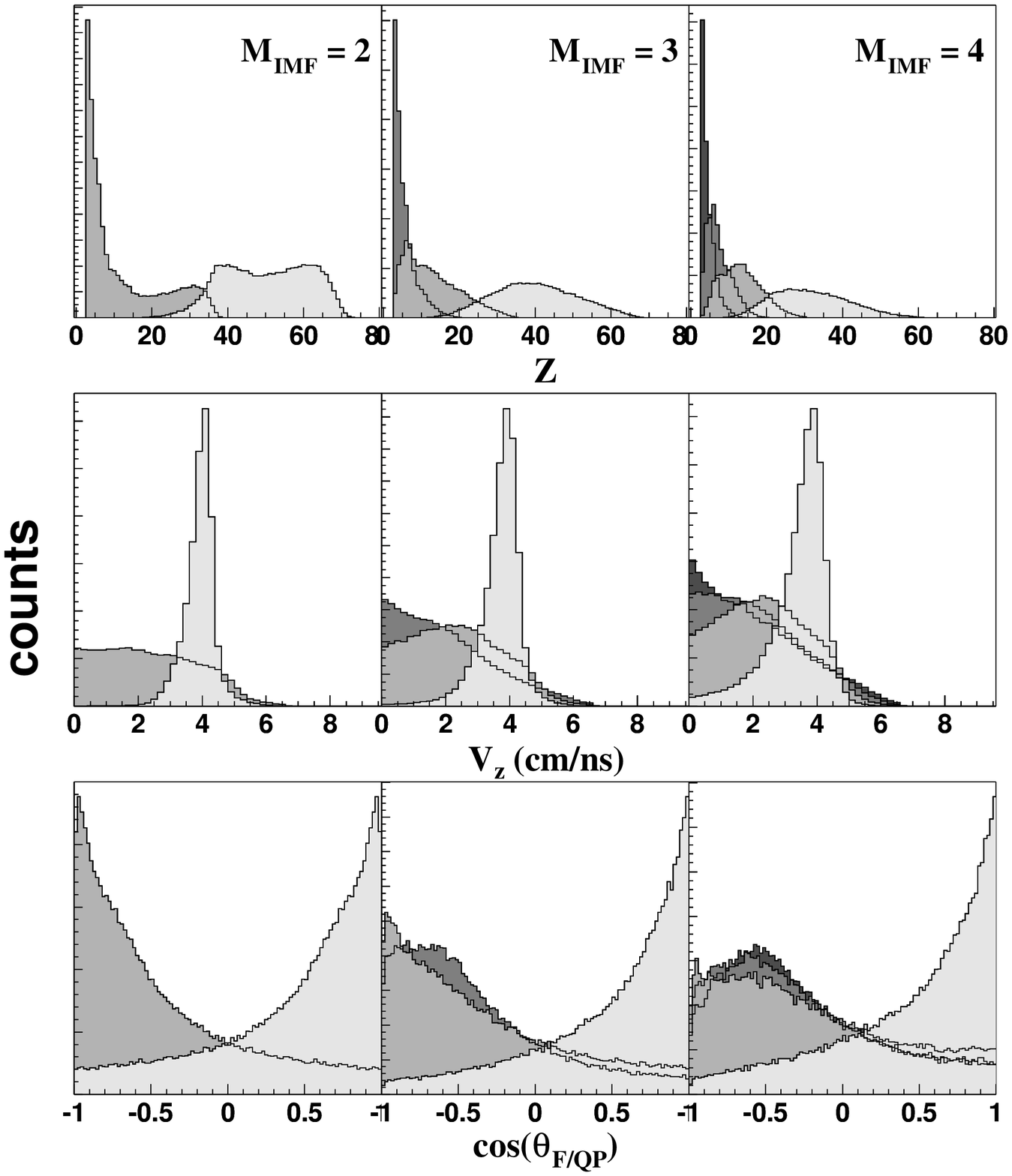}{Same as \ref{f:hier_taau33} for the Ta+Au 
system at 39.6 \AMeV.}{f:hier_taau39}{0.37}
\figart{htb}{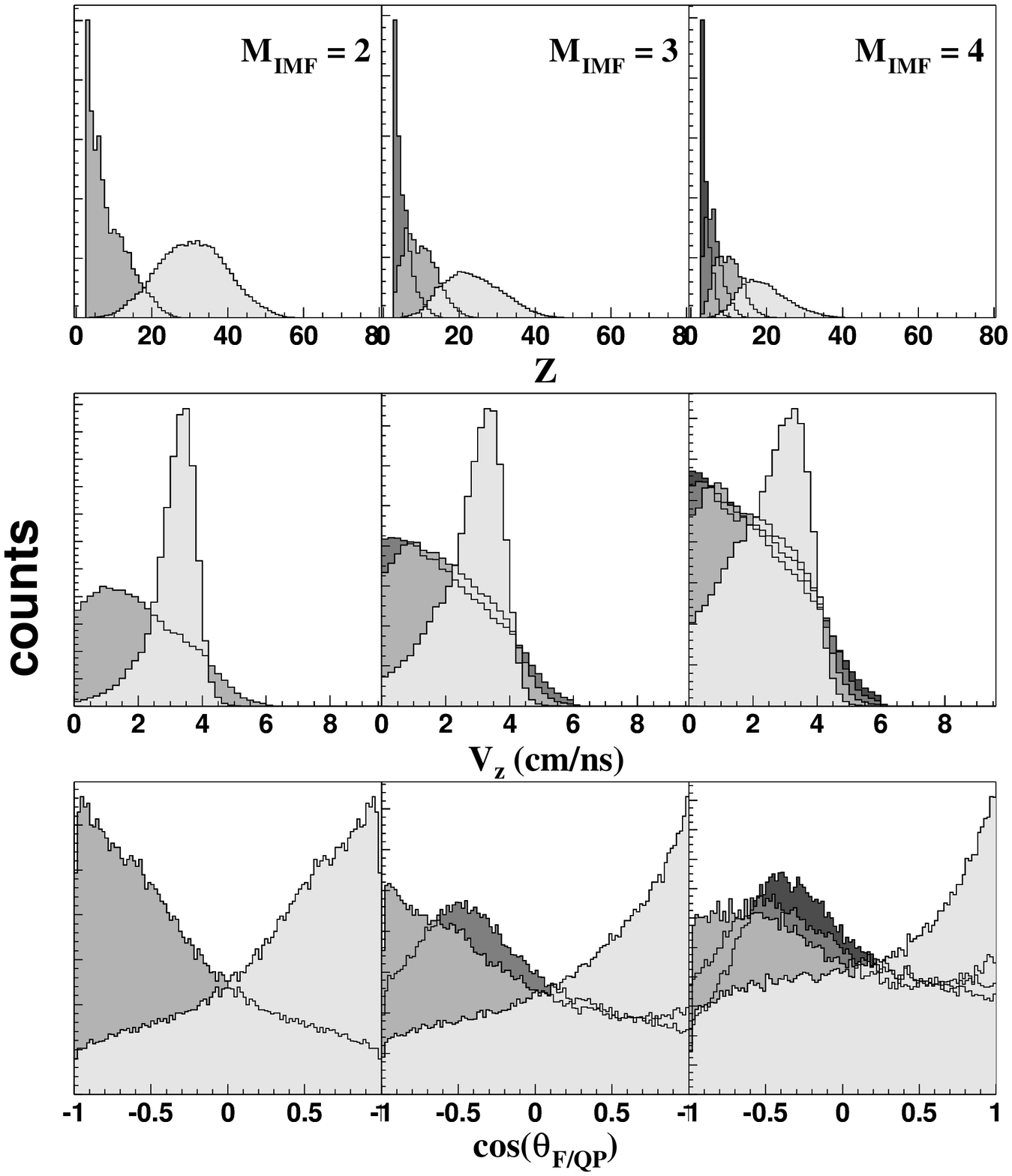}{Same as \ref{f:hier_taau33} for the Xe+Sn 
system at 39 \AMeV.}{f:hier_xesn39}{0.37}
\figart{htb}{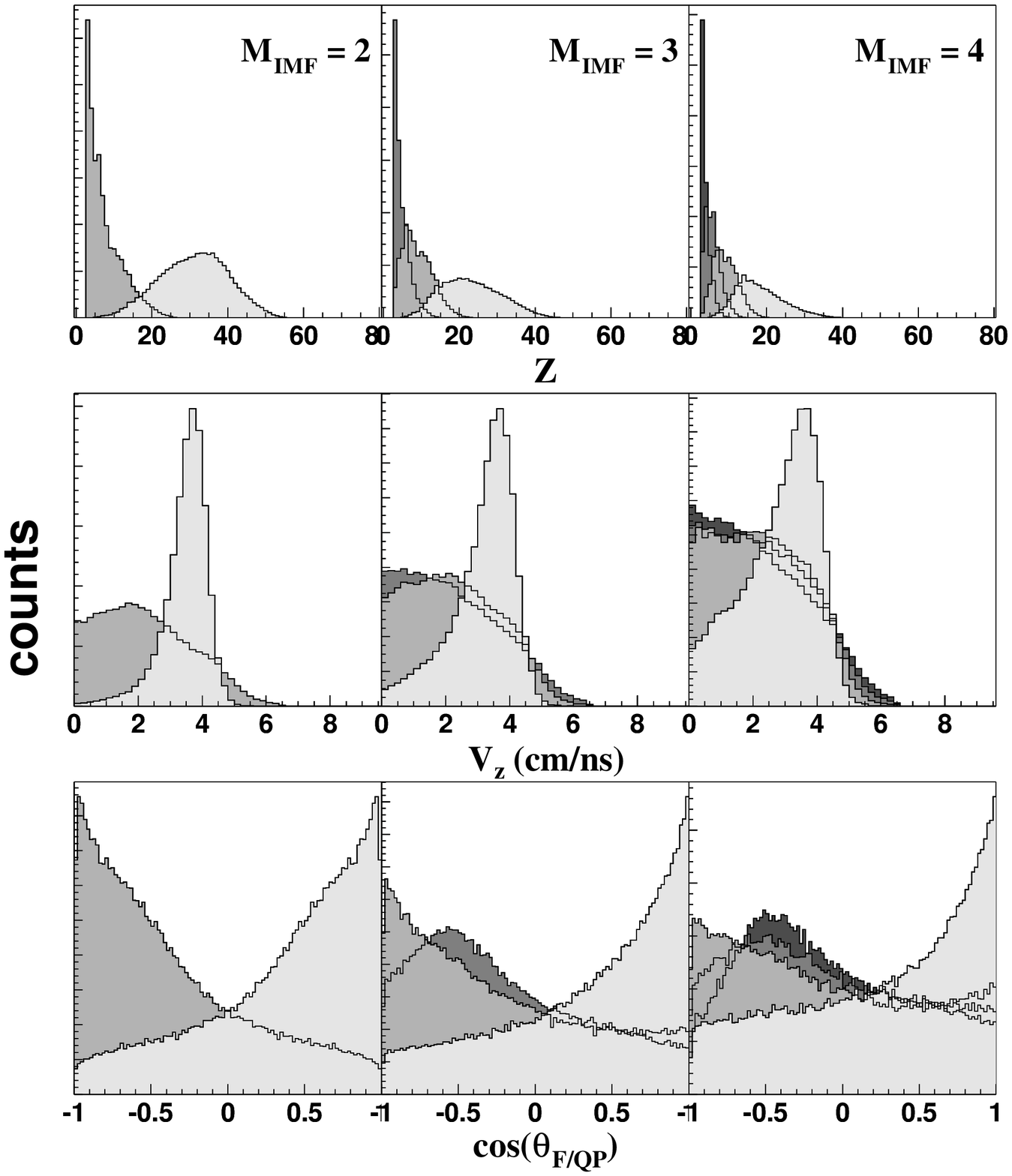}{Same as \ref{f:hier_taau33} for the Xe+Sn 
system at 45 \AMeV.}{f:hier_xesn45}{0.37}
\figart{htb}{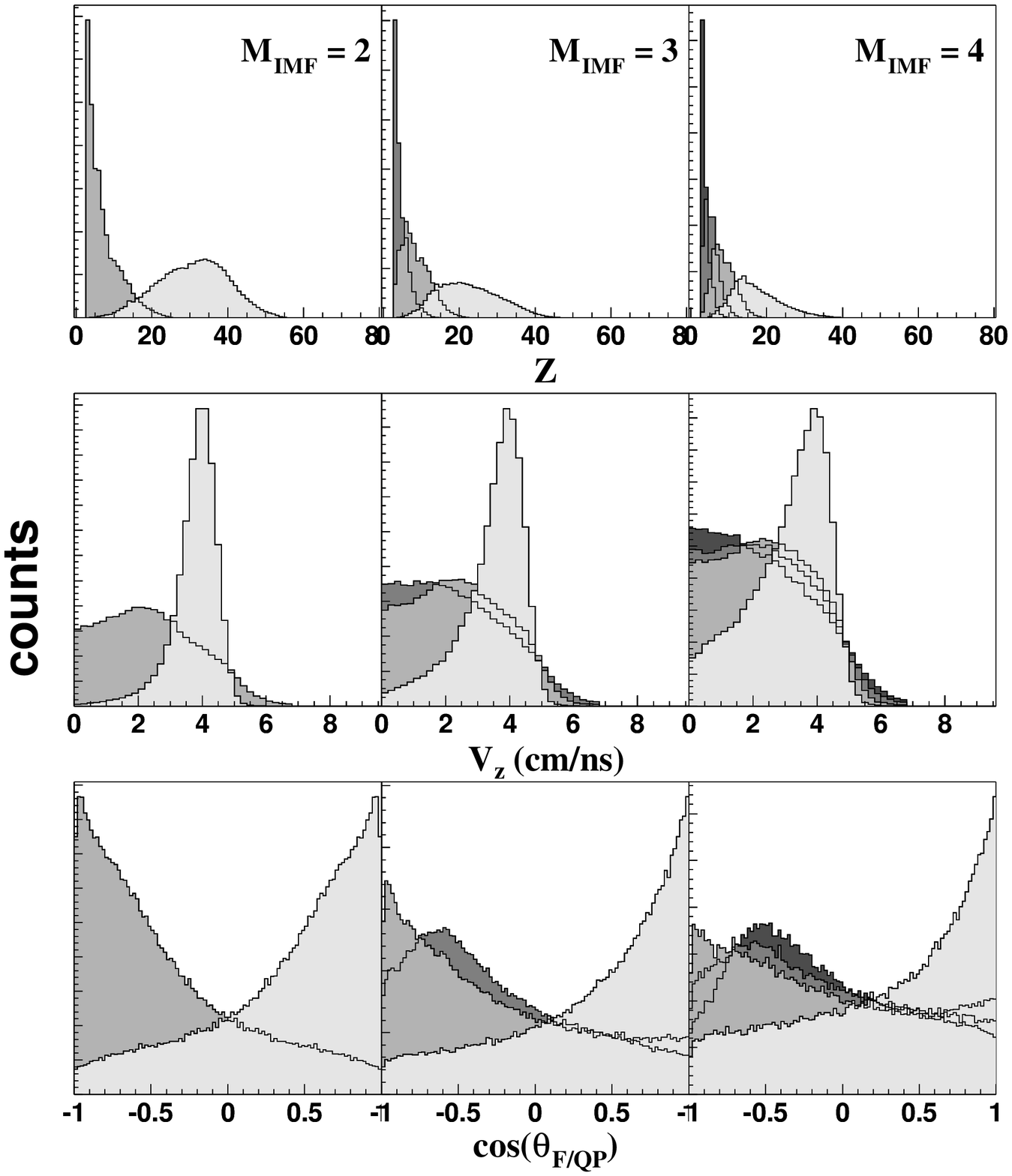}{Same as \ref{f:hier_taau33} for the Xe+Sn 
system at 50 \AMeV.}{f:hier_xesn50}{0.37}
\figart{htb}{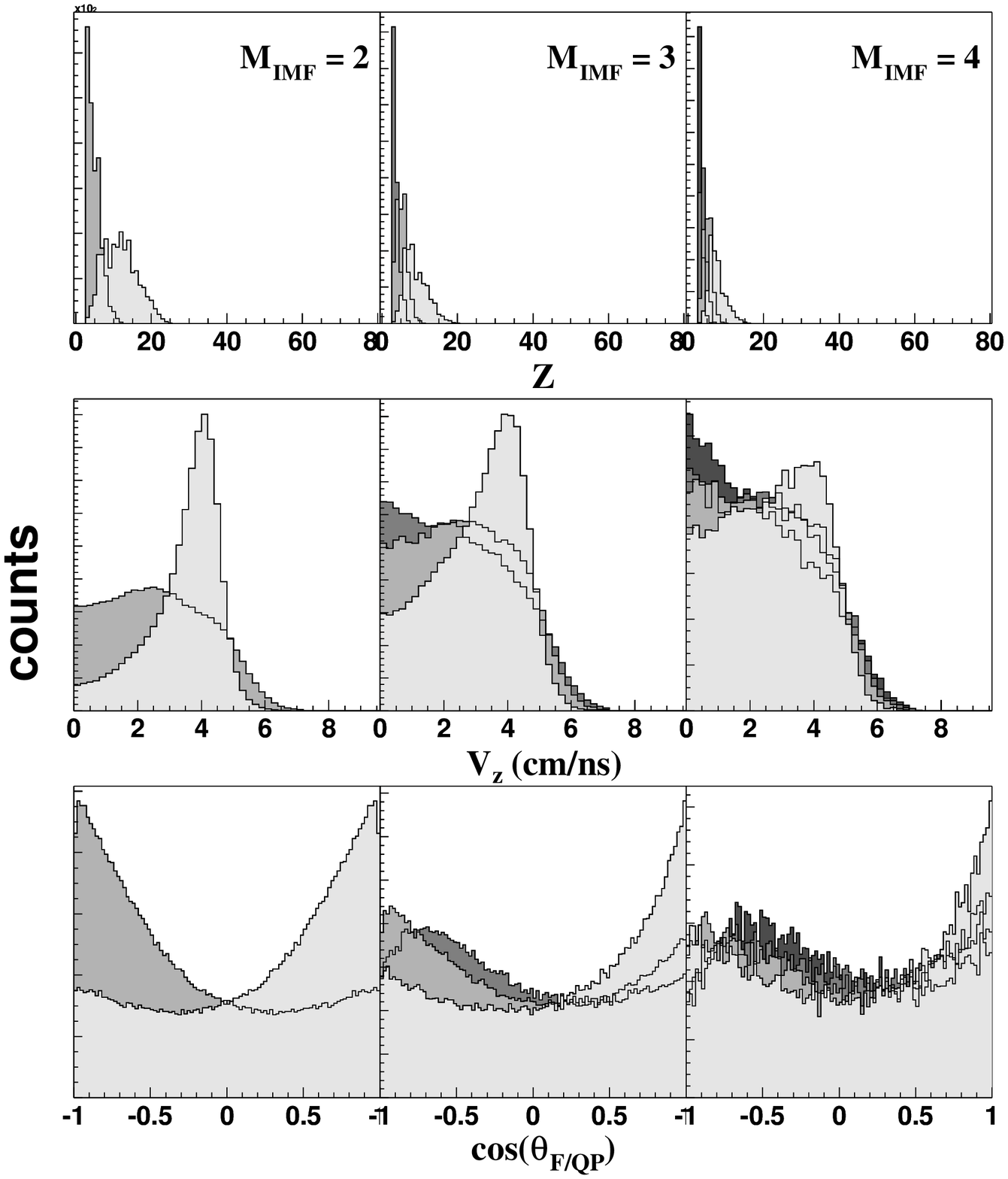}{Same as \ref{f:hier_taau33} for the Ni+Ni 
system at 52 \AMeV.}{f:hier_nini52}{0.37}
\figart{htb}{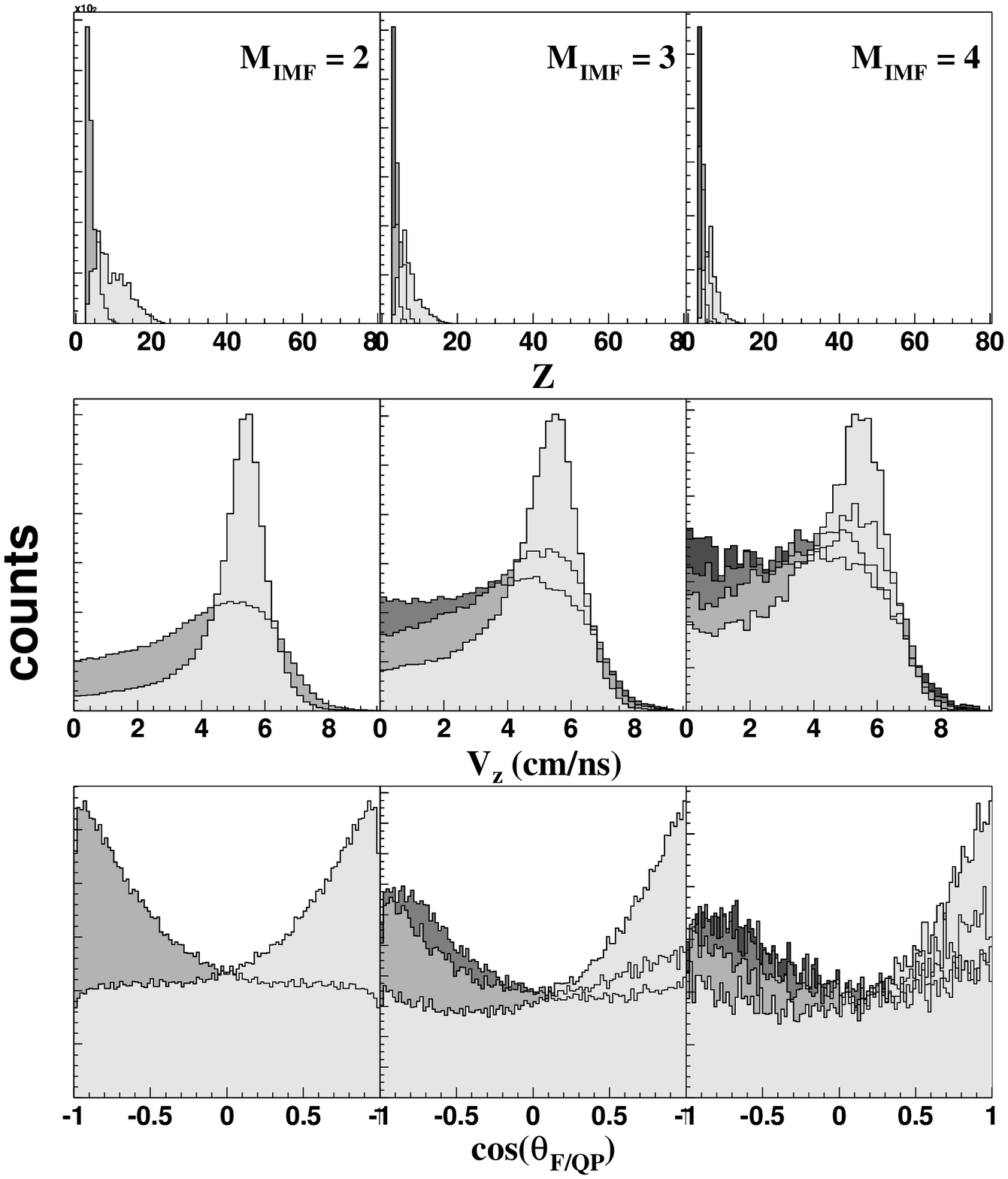}{Same as \ref{f:hier_taau33} for the Ni+Ni 
system at 90 \AMeV.}{f:hier_nini90}{0.37}
\figart{htb}{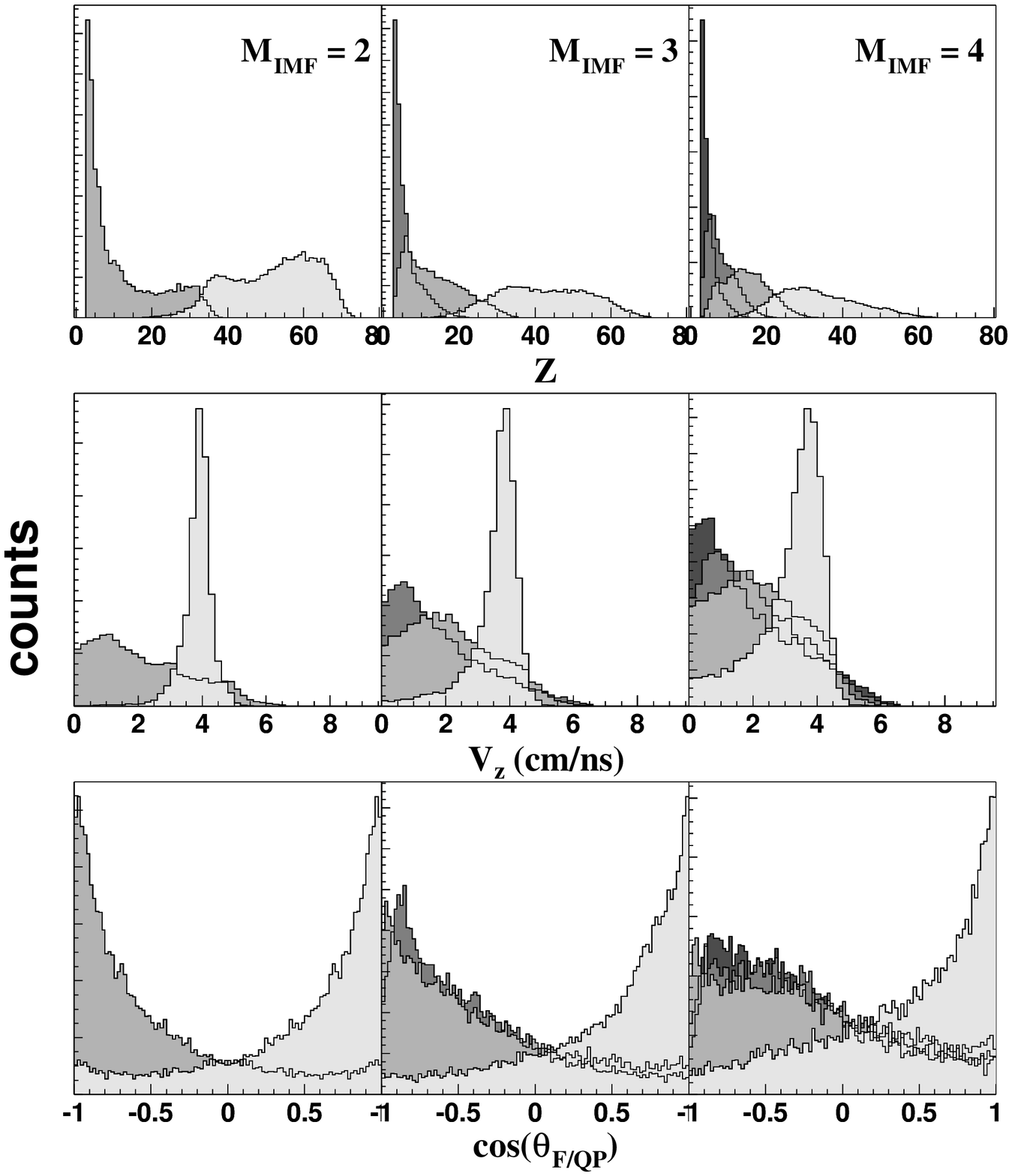}{Same as \ref{f:hier_taau33} for the Ta+U 
system at 33 \AMeV~($\theta_{QP} > 10^{\circ}$).}{f:hier_tau33}{0.37}
\figart{htb}{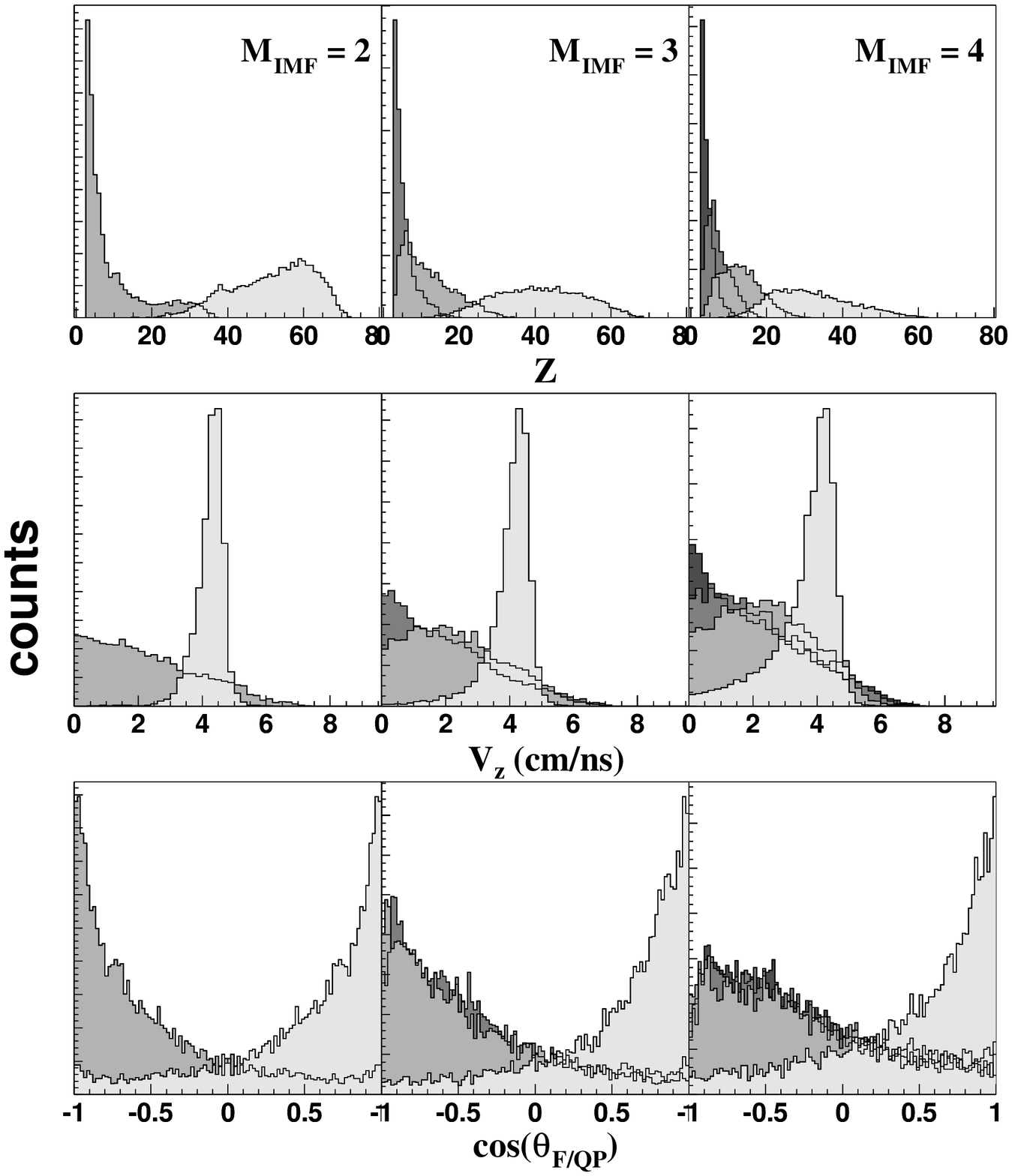}{Same as \ref{f:hier_taau33} for the Ta+U 
system at 39.6 \AMeV~($\theta_{QP} > 10^{\circ}$).}{f:hier_tau39}{0.37}
\figart{htb}{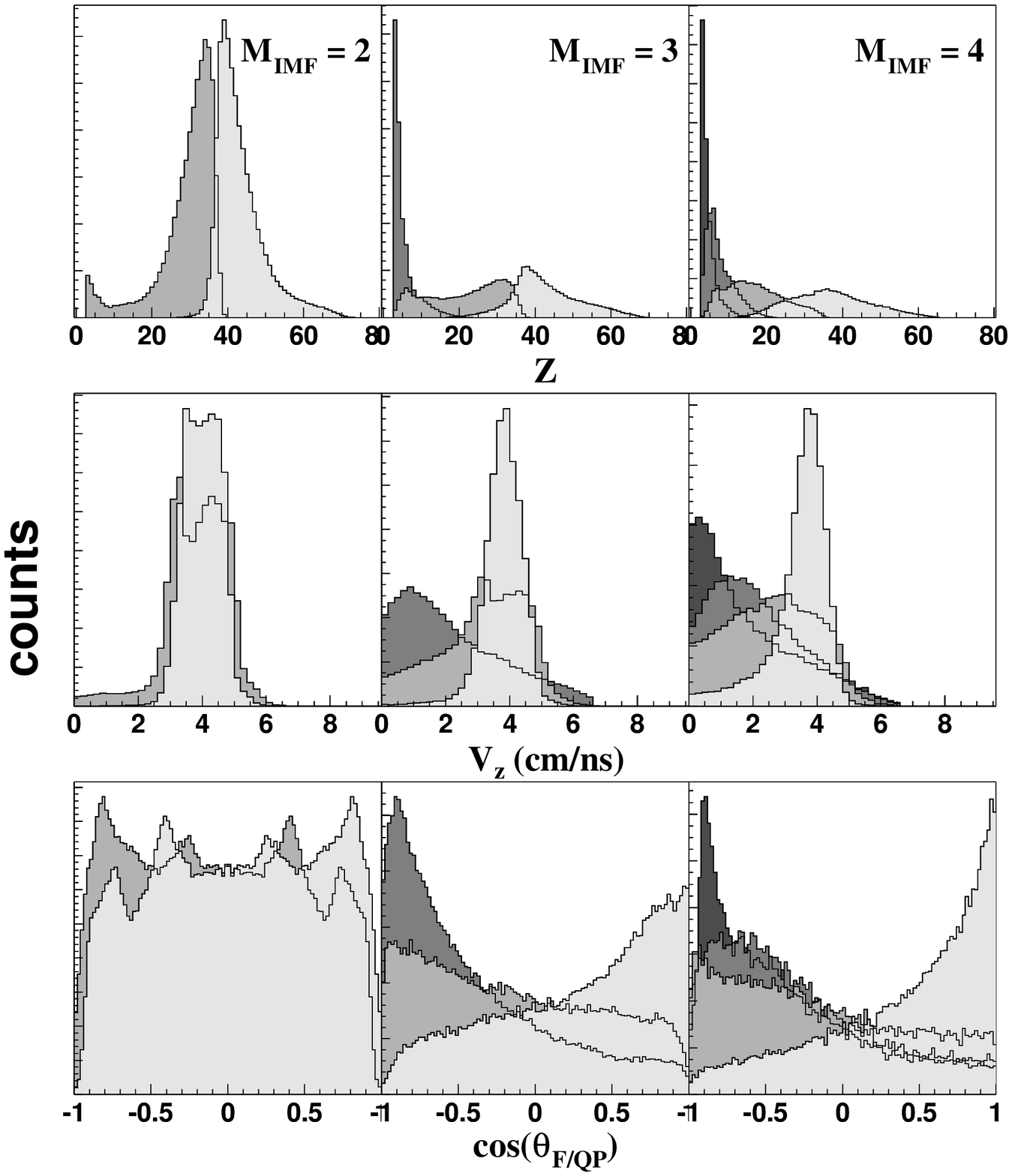}{Same as \ref{f:hier_taau33} for the Ta+C,U 
system at 33 \AMeV~(all $\theta_{QP}$).}{f:hier_tau33_all}{0.37}
\figart{htb}{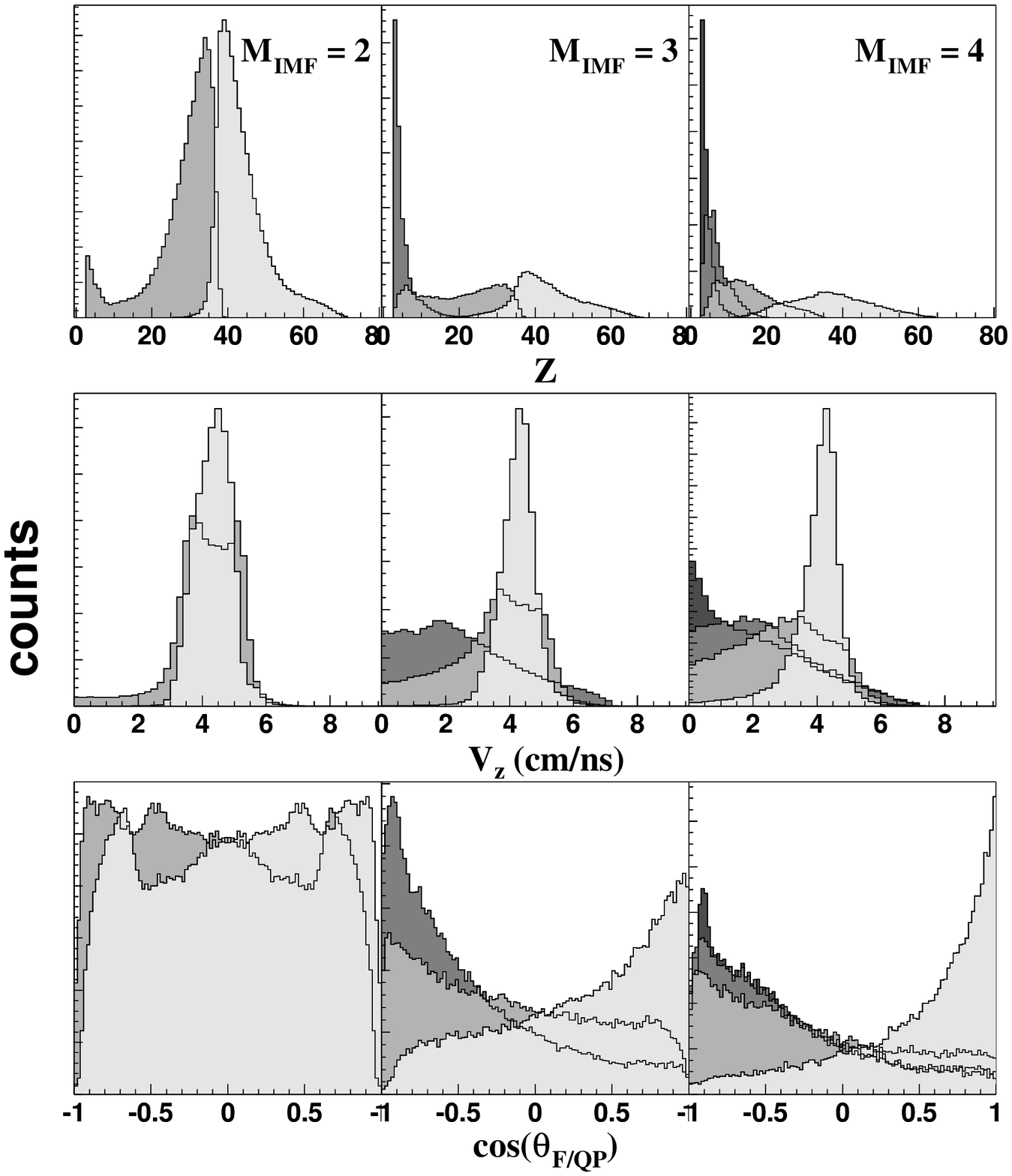}{Same as \ref{f:hier_taau33} for the Ta+C,U 
system at 39.6 \AMeV~(all $\theta_{QP}$).}{f:hier_tau39_all}{0.37}
\figart{htb}{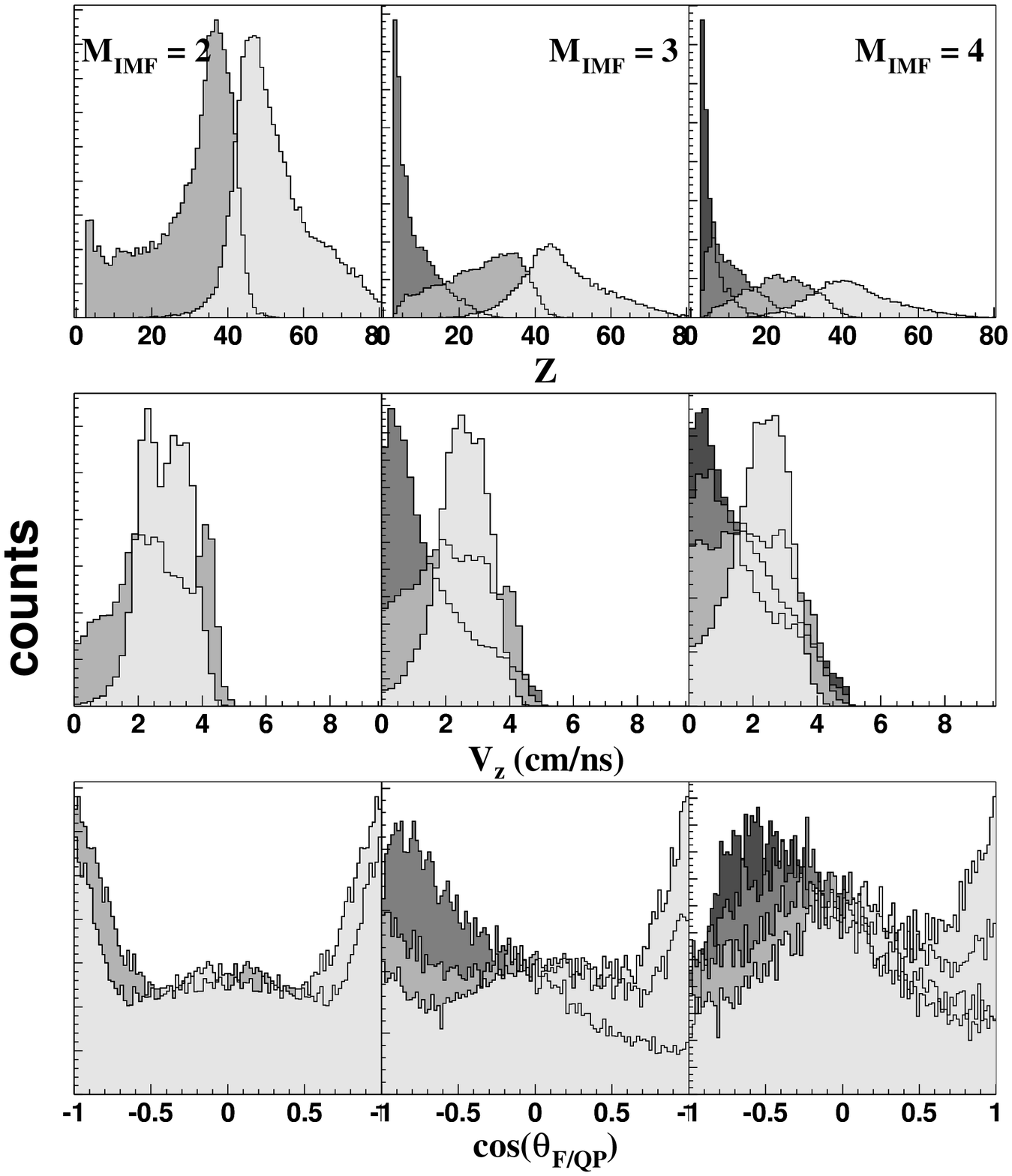}{Same as \ref{f:hier_taau33} for the U+U 
system at 24 \AMeV~($\theta_{QP} > 10^{\circ}$).}{f:hier_uu24}{0.37}
\figart{htb}{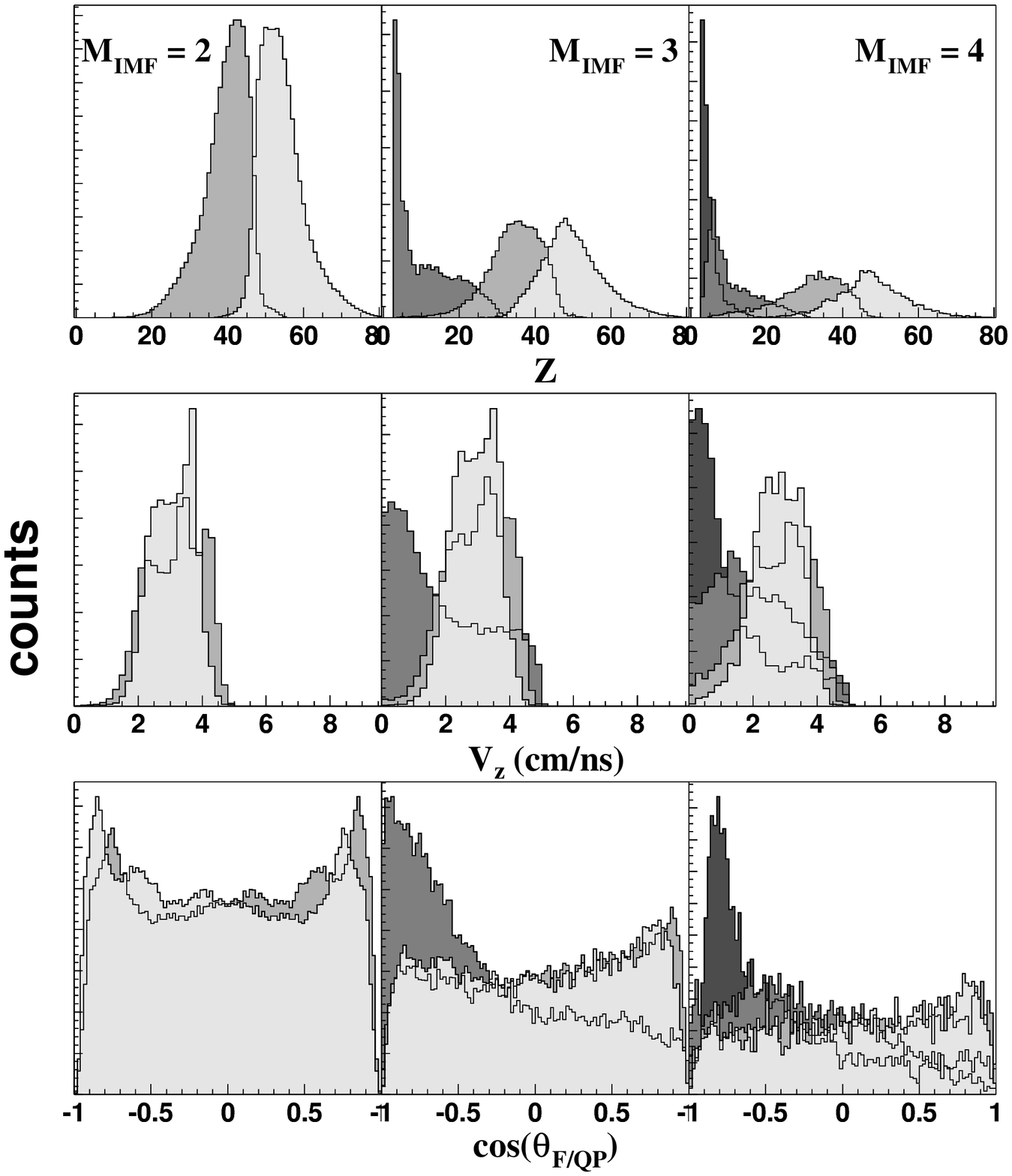}{Same as \ref{f:hier_taau33} for the U+C 
system at 24 \AMeV~($\theta_{QP} \leq 10^{\circ}$).}{f:hier_uc24}{0.37}
\figart{htb}{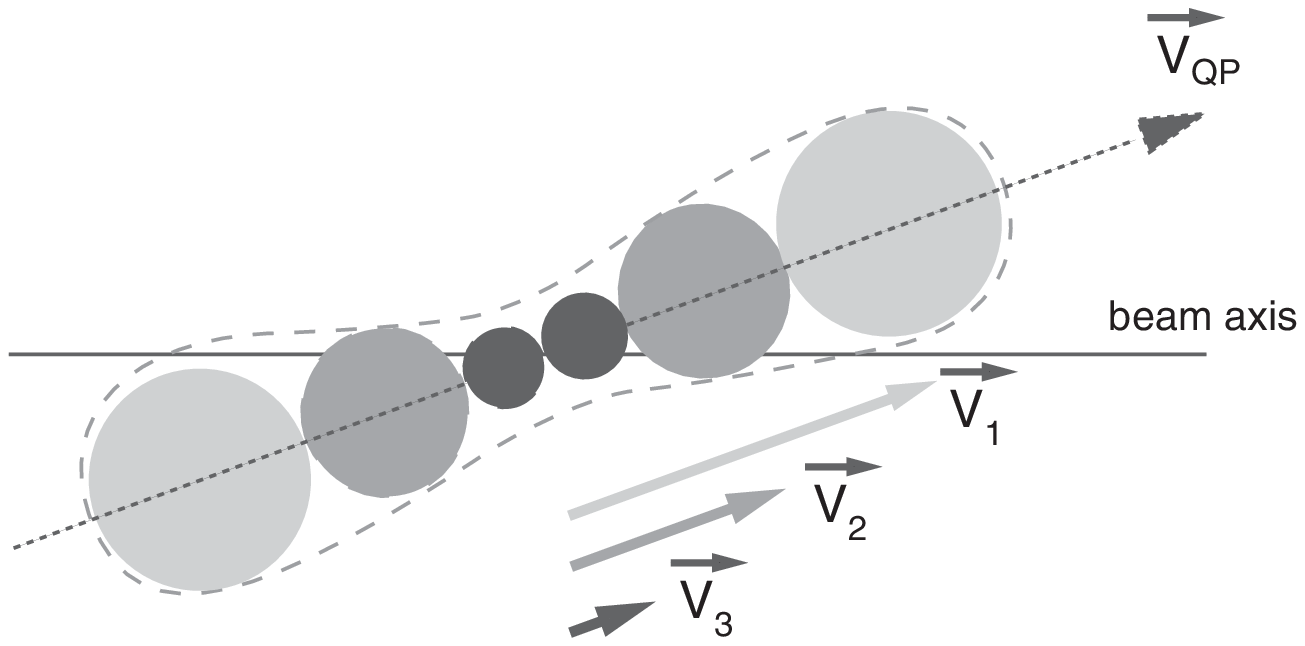}{Schematic view of the fragmentation scenario leading to
the ``hierarchy effect''. The shading darkens according to the charge ranking of
the fragments.}{f:schcol}{0.6}

\section{Size and velocity hierarchy}


We will first look at simple and direct  observables like the \modi{fragment} 
charge distributions, the \modi{corresponding 
distributions of the velocity component parallel to the beam ($V_{z}$)} and
the angular distributions for the different systems. 
\modi{The angle $\theta_{F/QP}$ is the angle between the velocity of
the fragment in the quasi-projectile (QP) frame and the velocity of 
the quasi-projectile 
in the center of mass
frame (see figure \ref{f:angles}). The events are sorted
according to the multiplicity, \Mfra, of the \mod{forward} fragments ($V_{z}
> V_{c.m.}$). In each event, the fragments are
sorted according to their charge.} 
 
\modi{Charge, \Vz~ and \ctfqp~} 
distributions are shown on figure
\ref{f:hier_taau33}. The uppermost row corresponds to the charge distribution,
the middle row to the \Vz~distribution and the lowermost row to \ctfqp
distribution. The
different columns correspond to the different fragment mutiplicities (from
\Mfra=2 to \Mfra=4 from left to right). On each \modi{panel, the shading 
of the
distribution darkens} according to the rank of the fragment in
the event.

One spectacular observation is that the velocity distributions \mod{are} 
strongly
correlated to the charge sorting:  the heaviest fragment \mod{is} 
in average the
fastest one, the second heaviest fragment is the second fastest one and so on.
For the \ctfqp~distributions, the distribution of the heaviest fragment is
peaked at forward angles ( \ctfqp$\approx 1$): the emission direction of this
fragment is close to the quasi-projectile velocity direction.  Such pictures
were not expected in case of a fragmentation of a fully equilibrated
quasi-projectile, for which the fragments are emitted in all directions without
a particular hierarchy for the velocities. This hierarchy effect (the
ranking in charge induces \mod{on average} 
the ranking in \Vz~ and the \ctfqp~ distribution of the
heaviest fragment is forward-peaked) suggests  an entrance channel effect.


\mod{Similar results are observed for the corresponding distributions for
most of other systems}. The heaviest fragment is the fastest one and its
direction of emission is close to the quasi-projectile velocity direction
whatever the system size and whatever the incident energy.  The hierarchy
between the size of the fragment and its velocity is observed for the Ta + Au
(figures \ref{f:hier_taau33} and \ref{f:hier_taau39} ), Xe+Sn (figures
\ref{f:hier_xesn39},  \ref{f:hier_xesn45} and \ref{f:hier_xesn50} ) and Ni+Ni
systems (figures \ref{f:hier_nini52}  and  \ref{f:hier_nini90}). This effect is
strong on the Ta+Au system and is slightly weaker for the Xe+Sn and the Ni+Ni
systems. \mod{For the Xe+Sn system, the anisotropy of the heaviest fragment
angular distribution increases slightly with the incident energy}.

For the Ta+U system (events with $\theta_{QP} > 10^\circ$, figures
\ref{f:hier_tau33} and \ref{f:hier_tau39} ), the same hierarchy \mod{effect} 
is observed as
for the Ta+Au system. For the Ta + C,U system (figures \ref{f:hier_tau33_all}
and \ref{f:hier_tau39_all} ), this hierarchy \mod{effect} is still observed for high
multiplicities (\Mfra$\geq 3$), but is not present for \Mfra=2. 
\modi{ This is due to the mixing between the
carbon backing and the uranium target mentioned in the previous section.} 
\modi{For \Mfra=2},
the two fragments have almost the same charges, the velocities are similar and
the whole range of \ctfqp is covered. The
contribution of the Ta+C collisions is stronger and stronger when
\Mfra~decreases. This contribution is especially seen for \Mfra=2  
(\mod{compare} leftmost
columns of figures \ref{f:hier_tau33} and \ref{f:hier_tau33_all} for 33 \AMeV~
and figures \ref{f:hier_tau39} and \ref{f:hier_tau39_all} for 39 \AMeV) where
the distributions are completely different. The Ta+C collisions lead to the
formation of an incomplete fusion nucleus, which decays through fission. In
this case, no hierarchy in velocities or no \modi{privileged angles} 
of the heaviest fragment \modi{are expected}.




For the U+U system (figure \ref{f:hier_uu24}, events with $\theta_{QP} >
10^\circ$) the hierarchy \mod{effect} is observed for \Mfra$\geq3$ but is not observed for
\Mfra=2. Due to the high fissility of uranium, the \Mfra=2 events
correspond mainly to the fission process. However, events for which the charge
of the heaviest fragment is close to 80 are still present. They correspond to
asymmetric binary break-ups which are not expected for the fission of such a
heavy nucleus. These asymmetric break-ups are not present for the U+C system
(figure \ref{f:hier_uc24}, leftmost column, \modi{$\theta_{QP} < 10^{\circ}$}) 
and no hierarchy effect is seen:
these observations are consistent with the fission of a fully equilibrated
nucleus. For \Mfra$\geq3$,  the velocities of the two heaviest fragments are
similar and their \ctfqp~distributions are flat. For the other fragments, the
hierarchy effect is present. This could be due to a two-step process in which
the hierarchy effect is present at the early stage of the collision, followed
by the fission of the heaviest fragment due to the high value of its charge.




To summarize this section, a hierarchy effect (the ranking in charge induces
the ranking in \Vz~ and the \ctfqp~ distribution of the heaviest fragment is
forward-peaked) is seen for many system sizes and incident energies. 
\mod{These} observations are not consistent with the
decay of a fully \mod{equilibrated} nucleus for which no hierarchy effect is 
expected.
This ``hierarchy effect'' is consistent with a strong deformation of 
\modi{QP or QT}
during the collision which is followed by the break-up of these elongated
\mod{nuclei} in two or more fragments (neck formation and break-up, \modi{see
figure \ref{f:schcol}}).
\mod{The fragments emitted by this neck reflect
its internal structure: its size at the center of mass 
is on average \modi{thinner than close to} the quasi-projectile or
quasi-target, and the velocity modulus in the center of mass frame 
of the nucleons in the neck close to the center of mass are smaller than 
\modi{those of nucleons close to the quasi-target or to the
quasi-projectile}. 

This leads to the ``hierarchy effect'' observed experimentally. 
\modi{The} velocity
gradient can be tracked by studying the relative velocities between the heaviest
fragment and the others. In the proposed scenario, this relative 
velocity is expected to
be larger than the one obtained \modi{from the decay of a 
fully equilibrated nucleus}.
For the heaviest systems, the standard fission process 
\modi{superimposes on} the neck formation and break-up. The two contributions 
are clearly seen for
Ta and U projectiles. In these cases, deviations from the standard 
\modi{fission} can be easily tracked.}

\section{Internal correlations}

In the previous section, \modi{we have evidenced a ``hierarchy effect''},
\modi{indicating a strong entrance channel influence on the}
fragmentation of the quasi-projectile. We will
study in this section correlations between some observables to \modi{verify
whether} the
fragmentation scenario proposed \modi{above resists a more} detailed
analysis. If an elongated quasi-projectile is formed and \modi{breaks quickly}, 
the
relative velocities between the fragments should be higher along the \modi{QP
velocity}
direction than in \modi{any} other direction, because the incident energy 
is not fully
damped in internal degrees of freedom in such a process. Within this
assumption, \modi{the heaviest fragment focused} along the QP velocity 
\modi{has a high velocity} because it is a remnant of the projectile.
In the case of the fragmentation of a fully \modi{equilibrated QP, 
the relative velocity should be independent of the break-up direction}. 

For \Mfra=2, the relative velocity between the two fragments can be
easily determined, \modi{while} it is not the case for higher 
multiplicities. We have
seen in the previous section that the heaviest fragment has the same properties
whatever the value of \Mfra: it is the fastest one and it is 
\mod{focused} in the
direction of the quasi-projectile \modi{in the center of mass frame}. 
We can use this property to define the
break-up direction $\theta_{prox}$ as the direction of the heaviest fragment
with respect to the quasi-projectile velocity, and the relative velocity 
$\overrightarrow{V_{rel}}$ as
the difference of the velocity of the heaviest fragment $\overrightarrow{V_{1}}$
and the velocity of the
center of mass of all other fragments detected with a velocity greater than
\Vcm. They are determined as follows:

\begin{equation}
\overrightarrow{V_{rel}}=\overrightarrow{V_{1}}-\frac{\sum_{i=2}^{M_{imf}}
Z_{i}\overrightarrow{V_{i}}}{\sum_{i=2}^{M_{imf}}Z_{i}} 
\end{equation} 

\noindent where $Z_{i}$ is the charge of the $i^{th}$ fragment ($Z_{i} >
Z_{i+1}$) and $\overrightarrow{V_{i}}$ its velocity

\begin{equation} \cos(\theta_{prox})=\frac{(\overrightarrow{V_{1}}
-\overrightarrow{V_{QP}}). \overrightarrow{V_{QP}}}
{|\overrightarrow{V_{1}}-\overrightarrow{V_{QP}}|
|\overrightarrow{V_{QP}}|} 
\end{equation} 

\noindent where 
$\overrightarrow{V_{QP}}$ is the velocity of the quasi-projectile. 
\modi{All the
velocities in formulae 1-3 are expressed in the center of 
mass frame.} \mod{With
this definition, $\theta_{prox}$ is identical to $\theta_{F/QP}$ of the heaviest
fragment (see figure \ref{f:angles}).}

\begin{equation} \overrightarrow{V_{QP}}=\frac{\sum_{i=1}^{M_{imf}}
Z_{i}\overrightarrow{V_{i}}}{\sum_{i=1}^{M_{imf}}Z_{i}} 
\end{equation} 

Let \modi{us} study the behaviour of the \ctp~ distributions. 
The experimental distributions are shown on figures 
\ref{f:asctp_taau33} and \ref{f:asctp_taau39} for the Ta + Au system at 33 and
39.6 \AMeV~respectively. On each figure, the columns correspond to different
values of \Mfra~ and the rows to different ranges in charge of the heaviest
detected fragment $Z_{1}$. This cut in $Z_{1}$ was done due to the particular
behavior of the heaviest fragment. It is also correlated to the asymmetry of
the break-up as in fission: the highest values of $Z_{1}$ correspond to the
more asymmetric break-ups (one big fragment and small other ones) and the
lowest values of $Z_{1}$ correspond to the most symmetric break-ups (equal size
fragments). It can be seen on these figures that the \ctp~ distributions are
\mod{in most cases} peaked at 1 
(emission along the direction of the QP velocity). 
\modi{The most forward peaked \ctp~ distributions are obtained for}
 the most asymmetric break-ups (high values of
$Z_{1}$) and this for all values of \Mfra. This is not expected in the case of
a fragmentation of a fully \modi{equilibrated} nucleus. \modi{An angular
momentum effect would lead to a forward-backward symmetry in angular
distributions}.

For the most symmetric break-ups (low values of $Z_{1}$) and the greater IMF
multiplicities, the distribution \modi{becomes less forward peaked}.
In this case,
the fragments having almost the same charge, \modi{the ranking in charge is
meaningless}. Consequently the memory of the direction
of the QP in the fragmentation process is blurred. Another possible scenario
could be the occurrence of the fragmentation of a fully 
\modi{equilibrated} QP. It will
be seen later that the cut \modi{at} 
\Vcm can also influence these distributions.   



\figart{htb}{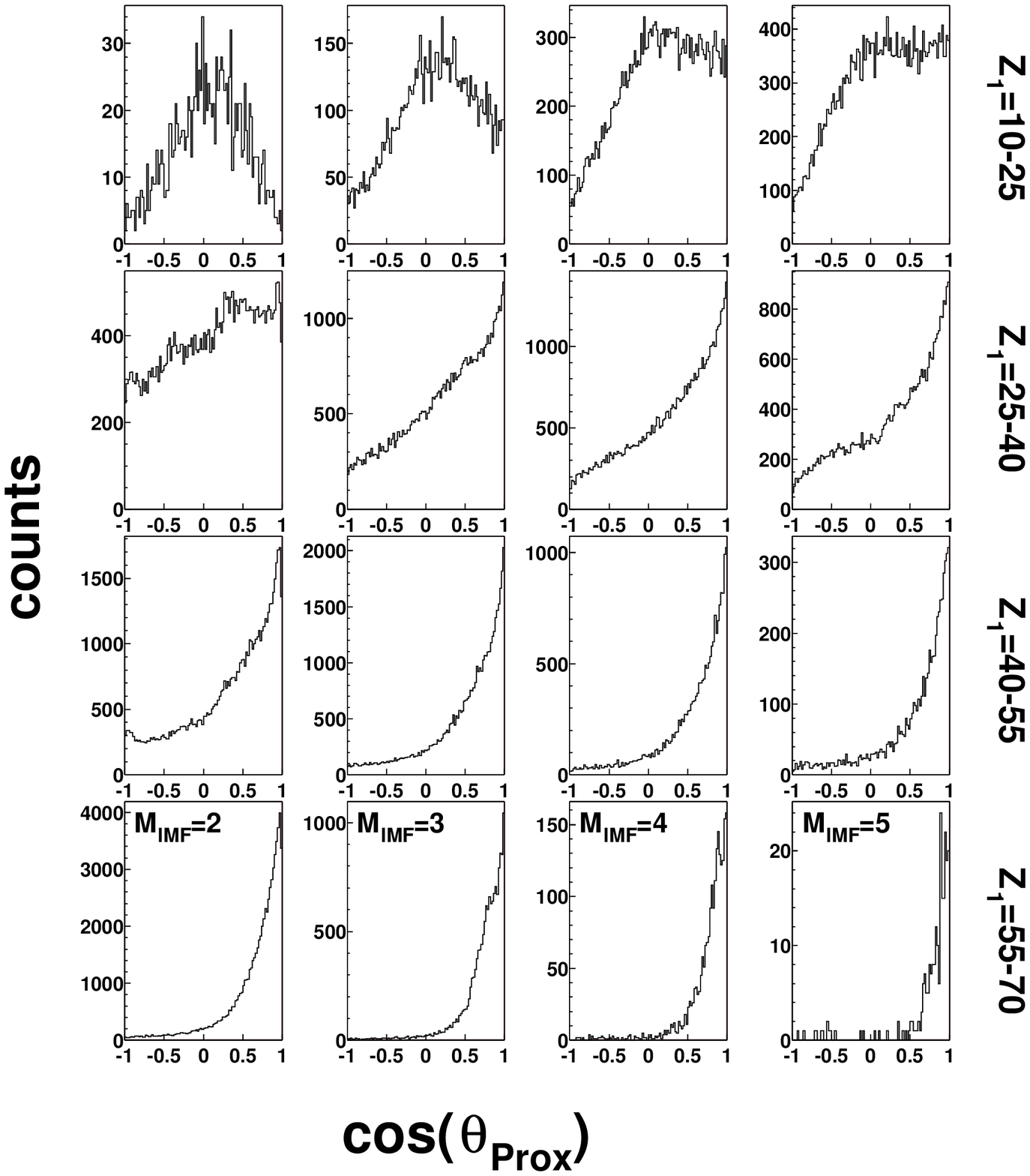}{\ctp~ distributions for the Ta+Au system at 33
\AMeV. The columns correspond to the different fragment multiplicities from 2
to 5. The rows correspond to different ranges for the charge $Z_{1}$  of the
heaviest fragment: the most symmetrical break-ups (low $Z_{1}$ values) correspond to
the uppermost row and the most asymmetrical break-ups 
(high $Z_{1}$ values) correspond to
the lowermost row.}{f:asctp_taau33}{0.45}
\figart{htb}{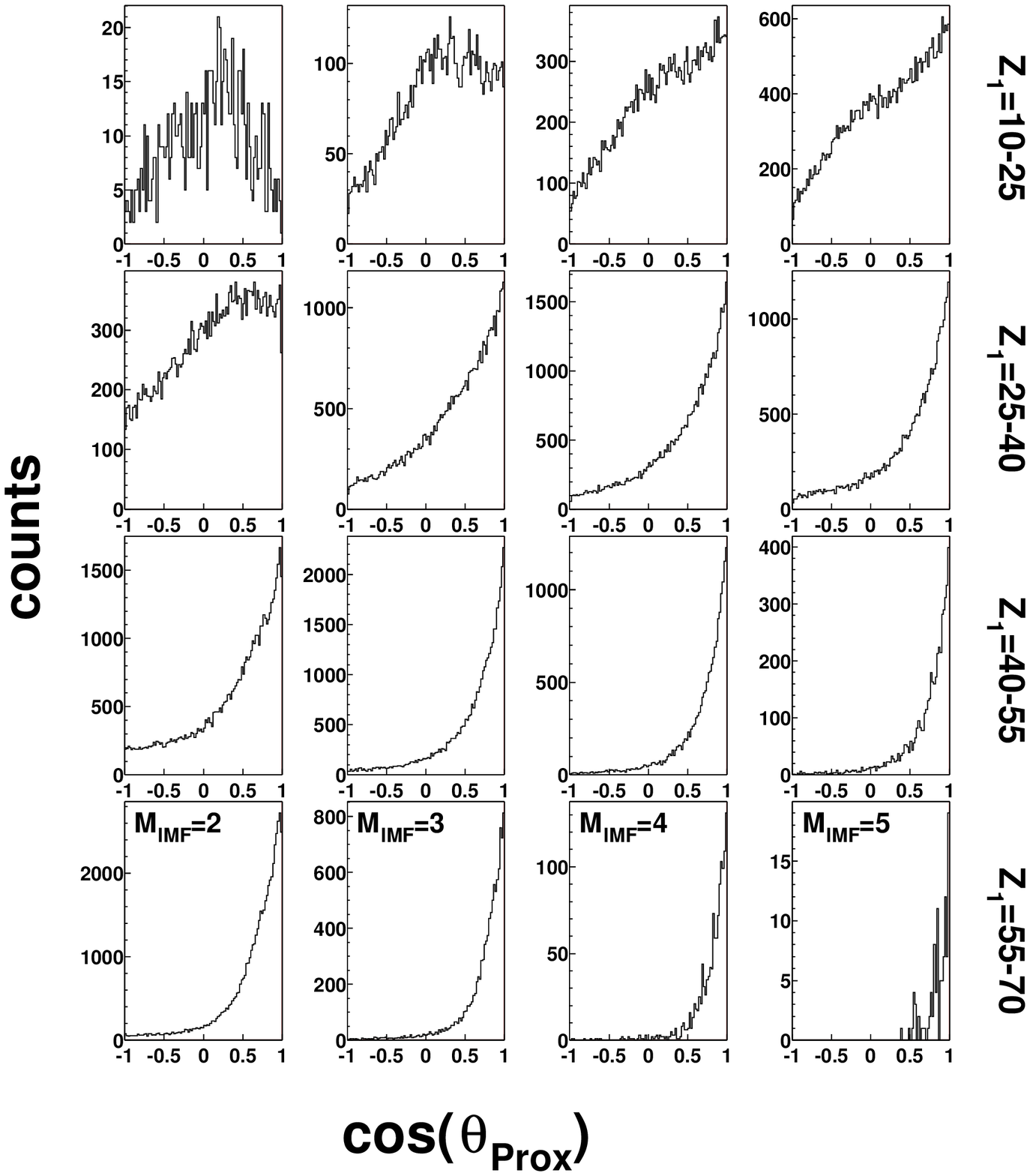}{Same as \ref{f:asctp_taau33} for the Ta+Au systeme
at 39.6 \AMeV.}{f:asctp_taau39}{0.45}



\modi{Is there a connexion between the strength of the anisotropy of the
\ctp~ distributions and the violence of the collisions?} 
The \modi{latter} has been estimated by using the sum of the
transverse energies \Etl~ of light charge particles ($Z \le 2$). This
global variable has been widely used as an impact parameter selector in
previous studies \cite{Plagnol2000,Bocage2000}. 
The low values of transverse energies correspond
to peripheral collisions and the high values to the most violent collisions.
The reduced  transverse energy distributions ($E_{t12}/E_{c.m.}$ where
$E_{c.m.}$ is the energy available in the center of mass) are presented in
figure  \ref{f:aset12_taau39} for Ta+Au collisions at 39.6 \AMeV. The shapes of
these distributions depend weakly on the value of \Mfra~ and \modi{strongly on
that of $Z_{1}$}. \modi{This is in agreement with $Z_1$ being the remnant
of the QP}. The average \Etl~ value decreases when going from
symmetrical (first row) to asymmetrical break-ups (last row) \modi{as already
shown for the lighter Ar+Ni system \cite{Dore2000a}}. The last row
corresponds to the most asymmetrical and aligned break-ups (see figure
\ref{f:asctp_taau39}). The transverse energies of the associated particles are
low and correspond to peripheral collisions. The first row corresponds to the
most symmetrical break-ups and to more wider  \ctp~ distributions.  The
transverse energies of the associated particles are high and correspond to more
central collisions.

It is \modi{interesting} to observe low \Etl~ values for the highest IMF
multiplicities \modi{(last row, rightmost panel of figure \ref{f:aset12_taau39})}. 
In this case, the excitation energy of the QP is low
and low \modi{evaporated} IMF multiplicities are expected. 
This observation, combined with the
``hierarchy effect'' and the strongly peaked \ctp~ distributions, strongly
suggests that these fragmentations are only due to entrance channel effects.
For the binary break-ups (first column), 
\modi{the very asymmetrical break-ups (last row)
are not expected for the standard fission of a quasi-Ta}. \modi{Nevertheless,
this is the dominant decay channel for \Mfra=2.}
The
second and  third \modi{panels of the first column (\Mfra=2)} 
correspond to more 
symmetrical break-ups ($Z_{1}\approx
37$) and the standard fission process is expected. In this case the
transverse \modi{energy distributions are} wider. 
It seems that two contributions could
be considered as already suggested by the  the \ctp~ distributions.
These observations are also made on the other systems.

\figart{htb}{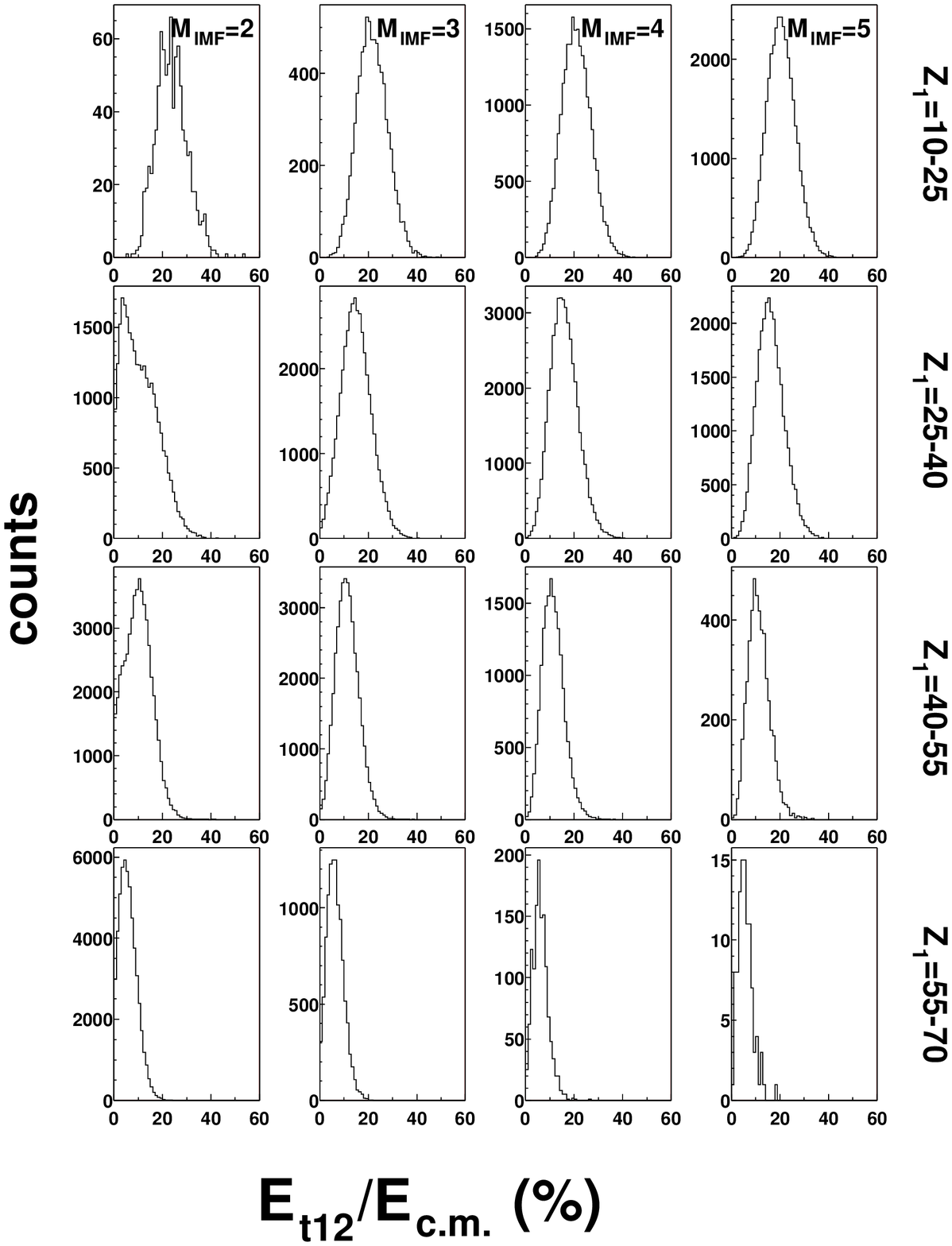}{$E_{t12}/E_{c.m.}$~ distributions (in percent) 
for the Ta+Au system at 39.6
\AMeV. The columns correspond to the different fragment multiplicities from 2
to 5. The rows correspond to different ranges for the charge $Z_{1}$  of the
heaviest fragment: the most symmetrical break-ups (low $Z_{1}$ values)
correspond to the uppermost row and the most asymmetrical break-ups  (high
$Z_{1}$ values) correspond to the lowermost row.}{f:aset12_taau39}{0.45}

\figart{htb}{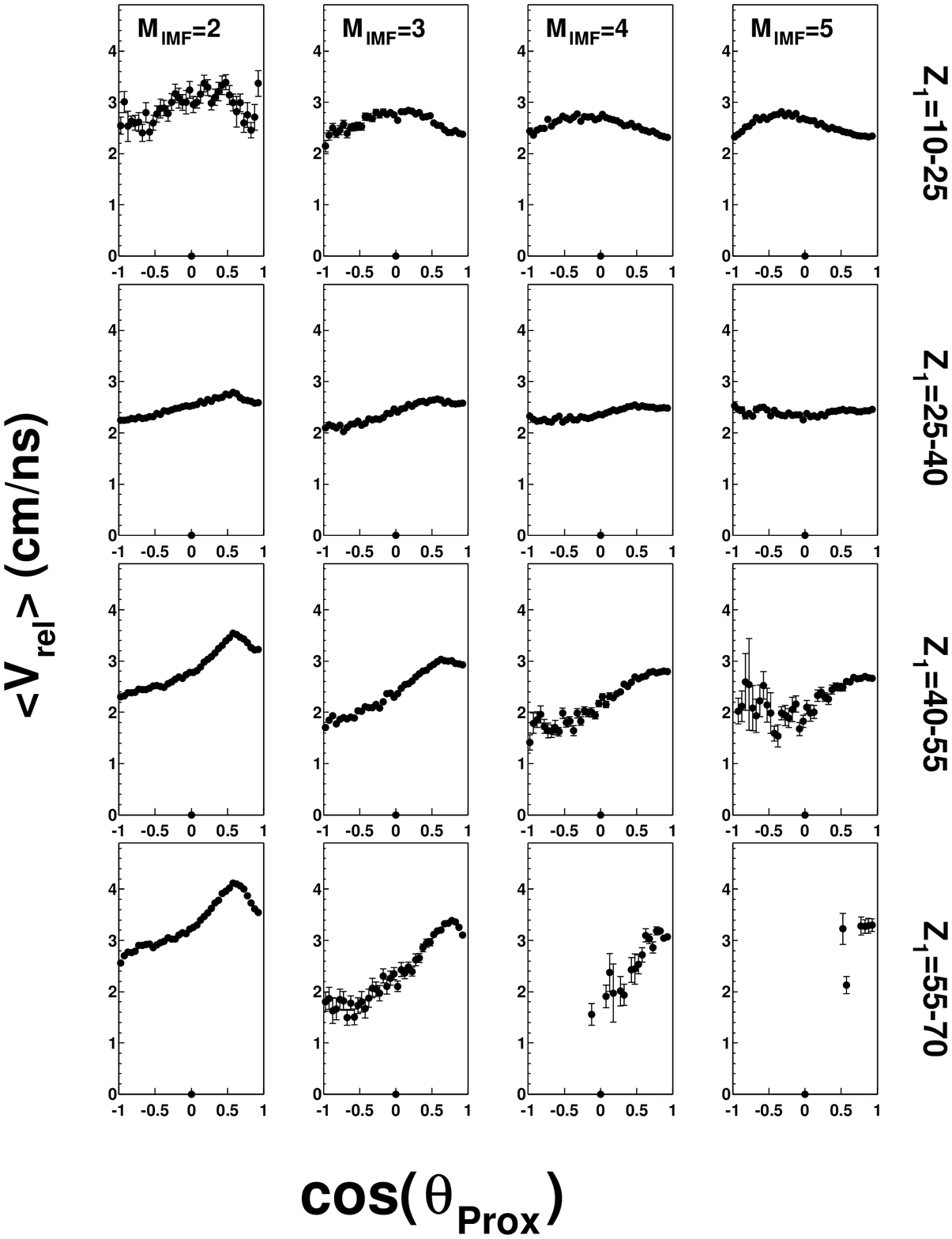}{Evolutions of the average relative velocity \Vr
with \ctp~ for the Ta+Au system at 39.6
\AMeV. The columns correspond to the different fragment multiplicities from 2
to 5. The rows correspond to different ranges for the charge $Z_{1}$  of the
heaviest fragment: the most symmetrical break-ups (low $Z_{1}$ values) correspond to
the uppermost row and the most asymmetrical break-ups 
(high $Z_{1}$ values) correspond to
the lowermost row.}{f:vrctp_taau39}{0.45}
\figart{htb}{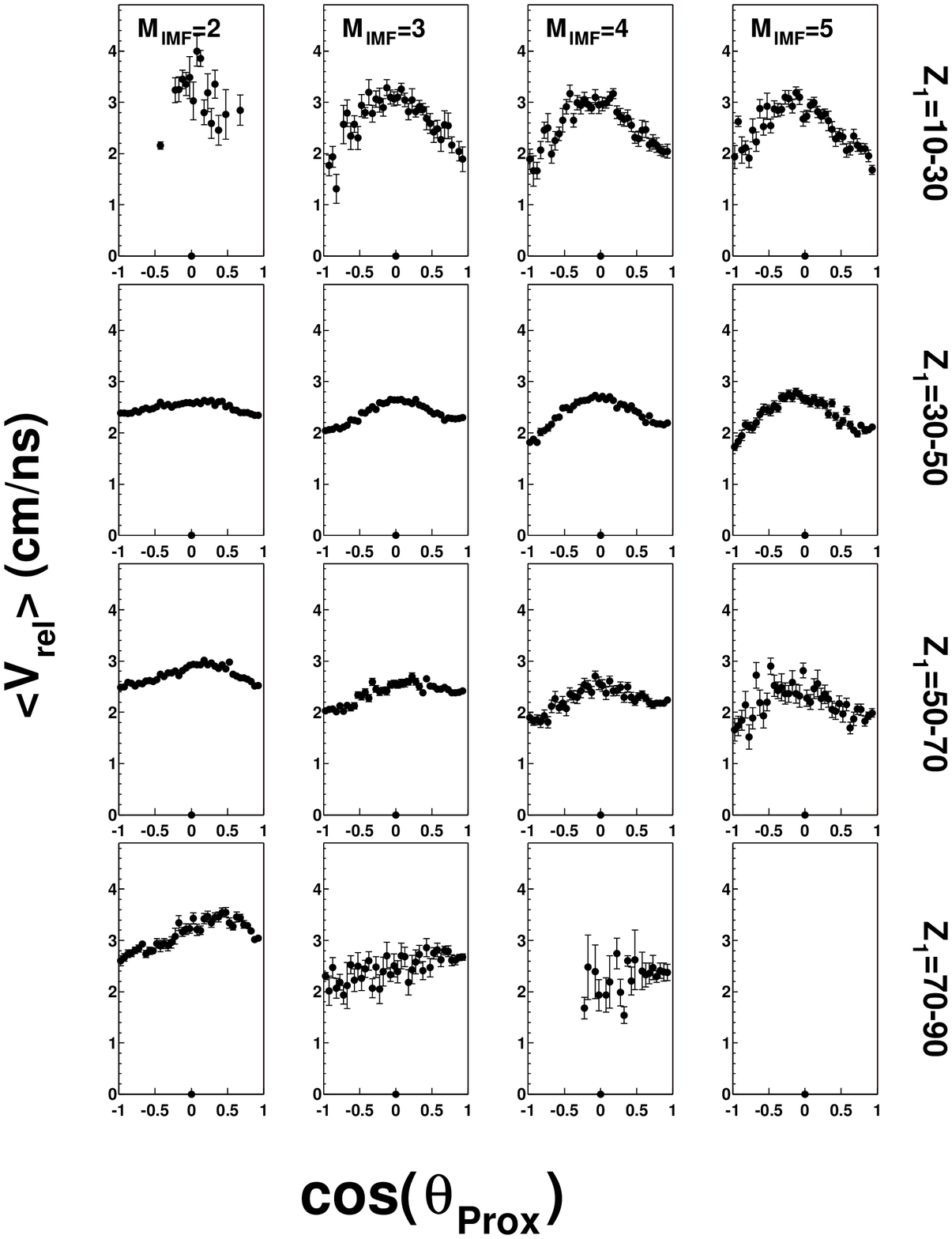}{Same as \ref{f:vrctp_taau39} 
for the U+U system at 24 \AMeV~($\theta_{QP} > 10^{\circ}$).}{f:vrctp_uu24}{0.45}
\figart{htb}{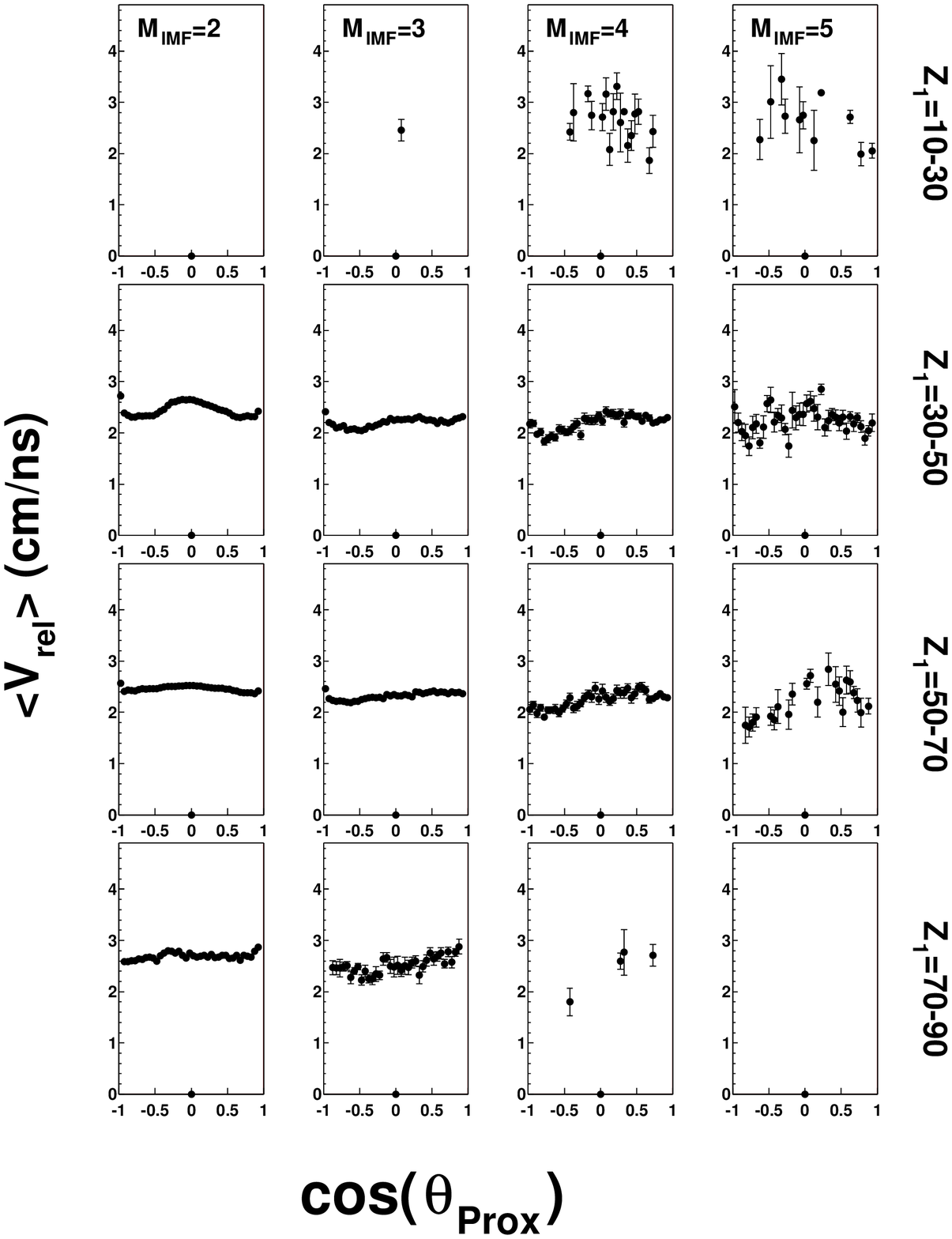}{Same as \ref{f:vrctp_taau39} 
for the U+C system at 24 \AMeV~($\theta_{QP} \leq 10^{\circ}$).}{f:vrctp_uc24}{0.45}
  
\figart{htb}{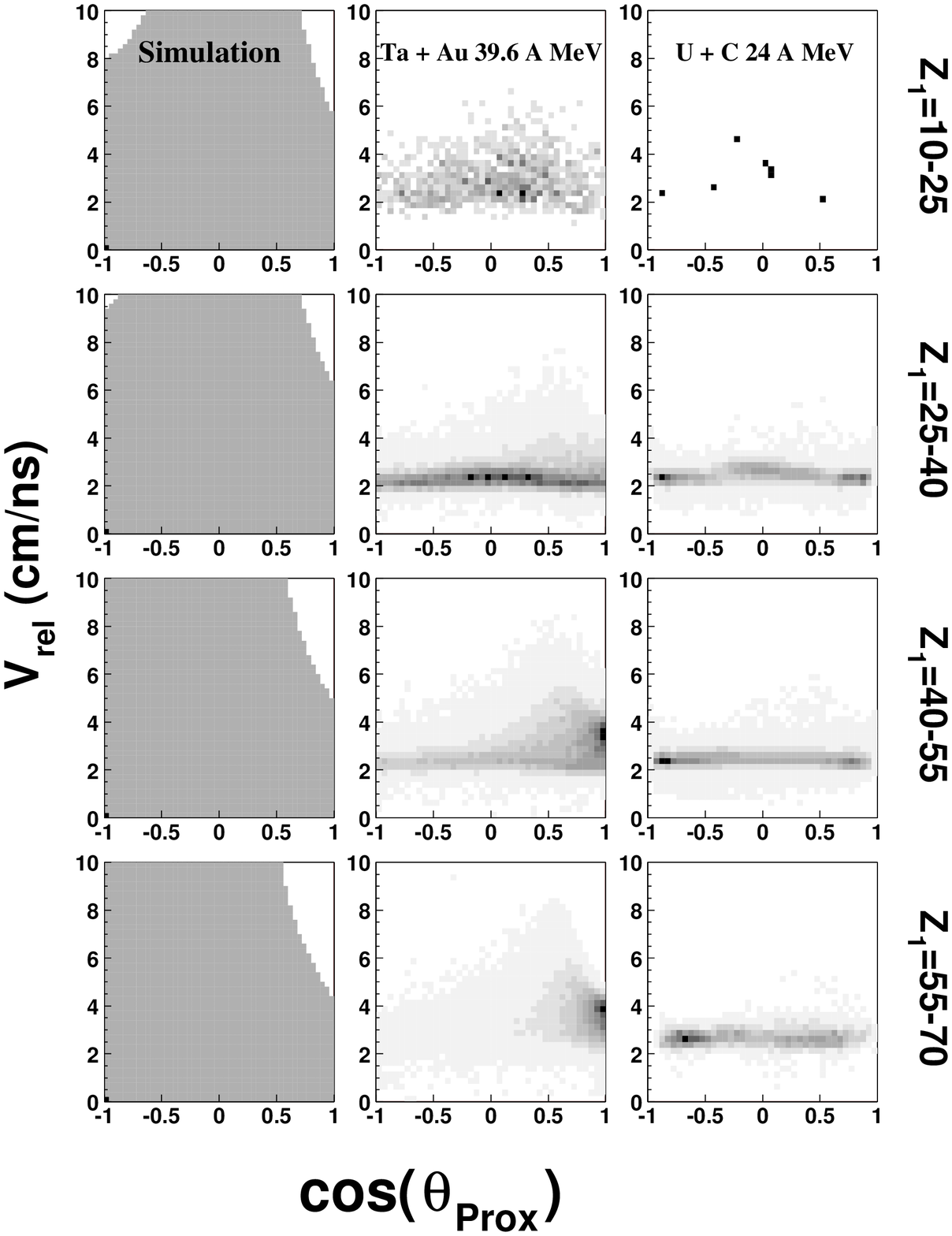}{First column: simulation of the accessible range of \Vr~
as a fonction of \ctp~ for the Ta+Au system at 39.6
\AMeV~and for \Mfra=2. The two other columns display the correlation between
\Vr~ and \ctp~ for two different systems. 
See text for more details.}{f:seuils_vrctp}{0.45}

Let us \modi{now come} to the correlation between the relative velocities of the
fragment \Vr~ and \ctp. They are presented on figure \ref{f:vrctp_taau39}  for
Ta+Au collisions at 39.6 \AMeV, on figure \ref{f:vrctp_uu24} for the  U+U
collisions at 24 \AMeV~and on figure \ref{f:vrctp_uc24} for the  U+C collisions
at 24 \AMeV. As in figures \ref{f:asctp_taau33} and \ref{f:asctp_taau39}, the
columns correspond to different IMF multiplicities and the rows to 
\modi{different}
values of $Z_{1}$. For the most asymmetric breakup of the Ta + Au collisions
(two lowest rows for figure \ref{f:vrctp_taau39}), a strong \mod{variation} 
of \Vr~ with \ctp~ is seen. The value of \Vr~ increases from 2.3 $cm/ns$  at 
\ctp=-1 up \modi{to} $\approx 3.5\quad cm/ns$ at \modi{around} 
$\cos(\theta_{prox})=0.8$ and then 
decreases. The
\mod{variation} is weaker for the more symmetric break-ups (two first rows). 
This
\mod{variation} is also observed for the U+U collisions but its amplitude is
smaller. For the U+C system (figure \ref{f:vrctp_uc24}), no \mod{evolution} 
is seen
whatever the IMF multiplicity and the charge of the heaviest fragment. For
the U+C system, no ``hierarchy effect'' was seen (see figure
\ref{f:hier_uc24}). 
\modi{Within} our interpretation 
\modi{of} the ``hierarchy effect'', 
no modulation is expected for this system. 
On \modi{all} other systems, the same correlation is observed: 
a modulation of \Vr~ with 
\ctp~ is seen when the ``hierarchy effect'' is observed. 
\modi{So the} modulation of \Vr~
with \ctp~ is correlated to the observation of a ``hierarchy effect''. 




The decrease of \Vr~ at the smallest angles (\ctp=1) is not compatible with
\mod{the expected behaviour of \Vr~ resulting from} our
interpretation on the ``hierarchy effect''. But due to the selection of the
fragments in this \modi{analysis} ($V_{z} > V_{c.m.}$),  
the highest values of \Vr~
can not be reached. We have \modi{tested} the effect of this velocity 
cut on our
\modi{analysis} with \modi{the} help of a simple simulation. 
For sake of simplicity, we \modi{have only
verified} the effect of the velocity cut for \Mfra=2. The basic 
\modi{ingredients}
of this simulation are the following: a QP with a charge $Z_{QP}$, a velocity
$V_{QP}$ and an angle $\theta_{QP}$ \modi{is} considered. 
This QP splits in two
fragments of charge $Z_{1}$ and $Z_{2}$, the axis of the break-up being
isotropically distributed and the relative velocity between 
\modi{the fragments} ranging from 0
cm/ns to 10 cm/ns. For each case, the event \mod{is taken into account in this
analysis} if
both fragments have a \Vz~ greater than \Vcm. The velocities
$\overrightarrow{V_{1}}$ and $\overrightarrow{V_{2}}$ of the two fragments are
calculated the following way:

\begin{equation}
\overrightarrow{V_{1}}=\overrightarrow{V_{QP}}+
(1-\frac{Z_{1}}{Z_{QP}})\overrightarrow{V_{rel}}
\end{equation}

\begin{equation}
\overrightarrow{V_{2}}=\overrightarrow{V_{QP}}-
\frac{Z_{1}}{Z_{QP}}\overrightarrow{V_{rel}}
\end{equation}
 
\mod{The first column \modi{of figure \ref{f:seuils_vrctp}} 
corresponds to the result of the simulation; the second
column \modi{shows the experimental} bidimensional \modi{plots} 
of \Vr~ versus \ctp~ for the Ta + Au
collisions at
39.6 \AMeV; the third column \modi{shows} 
the same plots for the U + C collisions at
24 \AMeV~\modi{($\theta_{QP} < 10^{\circ}$)}. 
\modi{The plots corresponding to events with an IMF multiplicity equal to 2 are
displayed.}
The rows \modi{correspond} to different selections on the charge of the
heaviest fragment $Z_{1}$, the uppermost row corresponding to the lowest $Z_{1}$
values and the lowermost row to the highest $Z_{1}$ values.}
For each
plot \mod{of the simulation (leftmost column)}, 
the values of  $V_{QP}$, $\theta_{QP}$ and 
the ratio $Z_{1}/Z_{QP}$ have
been chosen according to the average experimental values for the Ta+Au
\mod{collisions}
at 39.6 \AMeV~for a given IMF multiplicity and for a fixed $Z_{1}$ range. 
The shaded areas \mod{(leftmost column)}
correspond to the \mod{events selected in our analysis}. 
One can see that the velocity cut ($V_{z} > V_{c.m.}$) systematically removes
the events with the highest \Vr~ values when the fragmentation axis is aligned
on the QP velocity direction. For \ctp=1, this effect is
always present. This could explain the decrease of \Vr~ for high \ctp~ values
observed in figures \ref{f:vrctp_taau39} and \ref{f:vrctp_uu24}. \mod{This is
confirmed by the \modi{bidimensional}
plot (second column) which was used to obtain figure \ref{f:vrctp_taau39}}. 
The effect of
the velocity cut is clearly seen for the highest $Z_{1}$ values. The
agreement between data and simulation is \mod{good}, showing that the velocity
cut limits the range of accessible \Vr~ values. 
\modi{For higher \Mfra~ values, similar conclusions are obtained.} 
Another interesting observation is that the \Vr~
distribution for \ctp$\approx -1$ is narrow, its average value 
\Vr$=2.3\quad cm/ns$
corresponds to the Viola systematics \cite{Viola} and is not affected by the
velocity cut at \Vcm. When \ctp~ increases, the width of the \Vr~ distribution
and its average value increase while the velocity cut has no influence, i.e.
for \ctp~ values below 0.8. Above \ctp=0.8, high values of \Vr~ cannot be
reached and then its average value decreases. For the U+C system \mod{rightmost
column}, it can be
seen that the \Vr~ distributions are always
narrow and centered around the Viola systematics value whatever the \ctp~
value. \mod{For these collisions, 
our experimental cut does not affect the evolutions of \Vr~ with
\ctp.}


The observations made in this section confirm the scenario proposed in the
previous section. For the U+C system, all observations are compatible with the
decay of a fully equilibrated nucleus: no modulation of \Vr~ with \ctp~ is seen
while at the same time the ``hierarchy effect'' is absent or very weak. For the
systems with a heavy target, the \mod{variation} of \Vr~ with \ctp~ 
is observed. This
indicates that a strongly deformed QP or a neck of matter is formed at the early
stage of the collision, and \modi{breaks} quickly into several fragments. 
In such a
process, the incident kinetic energy is not fully damped, and the 
fragments keep the memory of the formation of the neck and/or deformed QP.
Within this interpretation, the heaviest fragment is a remnant of the
projectile. This scenario seems to occur for a wide range of system size and
incident energies. For the heaviest systems, like U+U for which the fissility of
the QP is high, it competes with the fission process.




\section{Conclusions}

The study of the parallel velocity distributions of the fragments 
\modi{forward emitted} as a function
of their charge exhibits a ``hierarchy effect'': the ranking in charge induces
in average a ranking in the average parallel velocity, the heaviest fragment is
the fastest and is \modi{focused} in the forward direction relative to the QP recoil
velocity. This ``hierarchy effect'' is stronger when the size of the target is
large, \modi{when the fissility of the QP is limited ($Z < 80$)} and when 
the incident energy is high. This
effect is observed for all systems, whatever the multiplicity of fragments.
These observations are compatible with the formation of a neck of matter
in-between the quasi-projectile and the quasi-target. \modi{The neck may be or
not attached to the QP (or QT) remnant. 
In any case, its break-up is fast enough to keep the memory of the entrance 
channel.} 


This mechanism is predominant for asymmetric break-ups which mainly correspond
\modi{to peripheral} collisions. For the most peripheral collisions, the
asymmetry observed for the binary break-ups is not compatible with the result
of a standard fission, and for the highest multiplicities, the high 
\modi{focusing}
of  the heaviest fragment at forward angles is inconsistent with the
fragmentation of a fully \modi{equilibrated nucleus}.

The observed \mod{variation} of the fragments relative velocity with the 
emission
\modi{angle} of the heaviest fragment strenghtens the previous statement. 
This
internal correlation study is a powerful tool to clearly establish the degree
of \modi{equilibration} of the studied ensemble. Up to now, the comparisons of 
models
to data were mainly made on global observations only (size distributions,
multiplicity distributions, kinetic energy distributions, and so on...). The
study of such internal correlations is a more accurate tool to \modi{test} the
pertinence of the models. 

In order to have a full understanding of the fragmentation process, 
\mod{the} models have to reproduce the effects of the entrance channel. A
first step could be to mock up these effects in effective parameters in the 
statistical decay approaches. \modi{It will be certainly worth to use dynamical
approaches \cite{BUU,LV,BLV,FMD,QMD,AMD,BNV,CNBD2002,Gingras2002} 
to test the presence of such a hierarchy effect}. But in all cases, 
the models have to reproduce
the global observations and the internal correlations as well. 




\end{document}